\documentclass[oneside,11pt]{article}

\usepackage[utf8]{inputenc}
\usepackage[T1]{fontenc}

\usepackage{amsthm}
\usepackage{amsmath}
\usepackage{amsfonts}
\usepackage{amssymb}

\usepackage{graphicx} 
\usepackage{float}
\usepackage{booktabs}
\usepackage{multirow}
\usepackage{rotating}
\usepackage{array}
\usepackage{xcolor}
\usepackage{subcaption}
\usepackage{algorithm}
\usepackage[noend]{algpseudocode}

\newcolumntype{R}[1]{>{\raggedleft\let\newline\\\arraybackslash\hspace{0pt}}m{#1}}
\newcolumntype{L}[1]{>{\raggedright\let\newline\\\arraybackslash\hspace{0pt}}m{#1}}

\usepackage{breakcites}

\usepackage{nowidow}

\usepackage{enumitem}

\usepackage[top=1.5cm,bottom=2cm,right=2.5cm,left=2.5cm]{geometry}
\usepackage{titling}

\usepackage{natbib}

\usepackage{ifplatform}

\usepackage{textcomp}

\ifwindows
\fi

\allowdisplaybreaks

\DeclareFontFamily{OT1}{pzc}{}
\DeclareFontShape{OT1}{pzc}{m}{it}{<-> s * [1.10] pzcmi7t}{}
\DeclareMathAlphabet{\mathpzc}{OT1}{pzc}{m}{it}

\newcommand{\CvM}{\text{CvM}}
\newcommand{\AD}{\text{AD}}

\makeatletter
\newcommand{\ssymbol}[1]{^{\@fnsymbol{#1}}}
\makeatother

\usepackage[pdftex,final,bookmarksnumbered,bookmarksopen=false,breaklinks,colorlinks]{hyperref}
\hypersetup{
	final,
	bookmarksnumbered=true,
	bookmarksopen=true,
	bookmarksopenlevel=0,
	unicode=false,          
	pdftoolbar=true,        
	pdfmenubar=true,        
	pdffitwindow=false,     
	pdftitle={Data-driven stabilizations of goodness-of-fit tests},    
	pdfdisplaydoctitle=true, 
	pdftoolbar=true,
	pdfmenubar=true,
	pdflang={English},
	pdfauthor={Alberto Fernandez-de-Marcos, Eduardo Garcia-Portugues},     
	pdfsubject={arXiv paper},   
	pdfcreator={Eduardo Garcia-Portugues},   
	pdfproducer={Eduardo Garcia-Portugues}, 
	pdfkeywords={Exact distribution}{Goodness-of-fit}{p-value}{Stabilization}{Uniformity}, 
	pdfnewwindow=true,      
	breaklinks=true,
	hidelinks,       
	linkcolor=black,          
	citecolor=black,        
	filecolor=magenta,      
	urlcolor=cyan           
}

\newcommand{\myref}[2]{\ref{#1}}%
\newcommand{\myeqref}[2]{\eqref{#1}}%

\newif\ifmain
\maintrue
\newif\ifsupplement
\supplementtrue

\newif\iffigstabs
\figstabstrue

\begin{document}

\ifmain

\title{Data-driven stabilizations of goodness-of-fit tests}
\setlength{\droptitle}{-1cm}
\predate{}%
\postdate{}%
\date{}

\author{Alberto Fern\'andez-de-Marcos$^{1,2}$ and Eduardo Garc\'ia-Portugu\'es$^{1}$}
\footnotetext[1]{Department of Statistics, Carlos III University of Madrid (Spain).}
\footnotetext[2]{Corresponding author. e-mail: \href{mailto:albertfe@est-econ.uc3m.es}{albertfe@est-econ.uc3m.es}.}
\maketitle

\begin{abstract}
Exact null distributions of goodness-of-fit test statistics are generally challenging to obtain in tractable forms. Practitioners are therefore usually obliged to rely on asymptotic null distributions or Monte Carlo methods, either in the form of a lookup table or carried out on demand, to apply a goodness-of-fit test. There exist simple and useful transformations of several classic goodness-of-fit test statistics that stabilize their exact-$n$ critical values for varying sample sizes $n$. However, detail on the accuracy of these and subsequent transformations in yielding exact $p$-values, or even deep understanding on the derivation of several transformations, is still scarce nowadays. The latter stabilization approach is explained and automated to (\textit{i}) expand its scope of applicability and (\textit{ii}) yield upper-tail exact $p$-values, as opposed to exact critical values for fixed significance levels. Improvements on the stabilization accuracy of the exact null distributions of the Kolmogorov--Smirnov, Cramér--von Mises, Anderson--Darling, Kuiper, and Watson test statistics are shown. In addition, a parameter-dependent exact-$n$ stabilization for several novel statistics for testing uniformity on the hypersphere of arbitrary dimension is provided. A data application in astronomy illustrates the benefits of the advocated stabilization for quickly analyzing small-to-moderate sequentially-measured samples.
\end{abstract}
\begin{flushleft}
\end{flushleft}

\section{Introduction}

The classical one-sample goodness-of-fit problem is concerned with testing the null hypothesis in which the cumulative distribution function (cdf) $F$ of an independent and identically distributed (iid) random sample $X_1, \ldots, X_n$ equals a certain prescribed cdf $F_0$. The most popular class of goodness-of-fit statistics for testing $\mathcal{H}_0:F=F_0$ is arguably that based on $F_n$, the empirical cumulative distribution function (ecdf) of $X_1, \ldots, X_n$. Ecdf-based test statistics confront $F_n$ against $F_0$, their best-known representatives being the Kolmogorov--Smirnov ($D_n$), Cramér--von Mises ($W_n^2$), and Anderson--Darling ($A_n^2$) statistics, all of them generating omnibus tests of $\mathcal{H}_0$ against $\mathcal{H}_1:F\neq F_0$. When $F_0$ is continuous, testing $\mathcal{H}_0$ reduces to testing whether the iid sample $U_1,\ldots,U_n$, $U_i := F_0(X_i)$, $i=1,\ldots,n$, is distributed as $\operatorname{Unif}(0, 1)$, the continuous uniform distribution on $(0,1)$. Hence, tests of uniformity, despite their a priori limited applicability, provide powerful approaches to most of the goodness-of-fit problems concerned with fully-specified null hypotheses. In particular, the above ecdf-based statistics have the attractive property of being distribution-free, i.e., their exact null distributions do not depend on $F_0$.

Both ecdf-based tests and uniformity tests have been exported to deal with data naturally arising in supports different from $\mathbb{R}$ or subsets thereof. This is the case of directional data, that is, data supported on the unit hypersphere $\mathbb{S}^{p-1}:=\{\mathbf{x}\in \mathbb{R}^{p}:\|\mathbf{x}\|= 1\}$, $p\geq2$, which commonly occurs in the form of circular ($p=2$) or spherical ($p=3$) data. The analysis of directional data faces specific challenges due to the non-Euclideanity of the support; see \cite{Mardia1999a} for a book-length treatment of tailored statistical methods and \cite{Pewsey2021} for a review of recent advances. In particular, tests of uniformity on $\mathbb{S}^{p-1}$ must be invariant under arbitrary rotations of the data coordinates, as these do not alter the uniform/non-uniform nature of the data. While a sizable number of tests of uniformity on $\mathbb{S}^{p-1}$ exist (see a review in \cite{Garcia-Portugues2020a}), perhaps the two most known omnibus tests are those of \cite{Kuiper1960} and \cite{Watson1961} on $\mathbb{S}^1$: their statistics, $V_n$ and $U_n^2$, can be regarded as the rotation-invariant versions of the Kolmogorov--Smirnov and Cramér--von Mises tests of uniformity, respectively. Moving beyond $\mathbb{S}^1$ has proven a challenging task for ecdf-based tests up to relatively recent years, with \cite{Cuesta-Albertos2009} using a Kolmogorov--Smirnov test on random projections data and \cite{Garcia-Portugues2020b} proposing a class of projected-ecdf statistics that extends \cite{Watson1961}'s test to $\mathbb{S}^{p-1}$ (see Section \ref{section:proj-based-unif-tests}). As in the classical setting, tests of uniformity on $\mathbb{S}^{p-1}$ allow for testing the goodness-of-fit of more general distributions: in $\mathbb{S}^1$, this is a straightforward application of the probability integral transform in the angles space $[-\pi,\pi)$; the case $\mathbb{S}^{p-1}$, $p\geq3$, is remarkably more complex and has been recently put forward in \cite{Jupp2020}.

\begin{table}[htbp!]
	\iffigstabs
	\small
	\centering
		\begin{tabular}{ >{\arraybackslash}m{1.5cm}  >{\arraybackslash}m{13.5cm}}
			\toprule
			Statistic & Exact distribution approximations\\
			\midrule
			$D_n$ & \cite{Massey1950, Massey1951}$\ssymbol{1}{}^{,}{}\ssymbol{2}$, \cite{Birnbaum1952}$\ssymbol{3}$, \cite{Maag1971}$\ssymbol{4}$, \cite{Marsaglia2003}$\ssymbol{3}$, \cite{Brown2007}$\ssymbol{2}{}^{,}{}\ssymbol{8}$, \cite{Facchinetti2009}$\ssymbol{8}$\\
			\midrule
			$W^2_n$ & \cite{Marshall1958}$\ssymbol{8}$, \cite{Pearson1962}$\ssymbol{5}{}^{,}{}\ssymbol{9}$, \cite{Tiku1965}$\ssymbol{3}$, \cite{Stephens1968}$\ssymbol{3}{}^{,}{}\ssymbol{5}{}^{,}{}\ssymbol{9}$, \cite{Knott1974}$\ssymbol{7}$, \cite{Csorgo1996}$\ssymbol{3}$\\
			\midrule
			$V_n$ & \cite{Stephens1965}$\ssymbol{1}$, \cite{Maag1971}$\ssymbol{4}$, \cite{Durbin1973, Arsham1988}$\ssymbol{8}$\\
			\midrule
			$U^2_n$ & \cite{Pearson1962}$\ssymbol{5}{}^{,}{}\ssymbol{9}$, \cite{Tiku1965}$\ssymbol{3}$, \cite{Quesenberry1977}$\ssymbol{9}$\\
			\midrule
			$A^2_n$ & \cite{Lewis1961}$\ssymbol{9}$, \cite{Marsaglia2004}$\ssymbol{6}$\\
			\bottomrule
	\end{tabular}
	\fi
	\caption{\small Summary of existing specific approaches for approximating exact distributions of several goodness-of-fit test statistics. The approximations rely of the following main techniques: difference equations$\ssymbol{1}$, recursive formulae$\ssymbol{2}$, truncated approximations$\ssymbol{3}$, asymptotic expansions$\ssymbol{4}$, approximation of distribution moments$\ssymbol{5}$, correction factors$\ssymbol{6}$, characteristic function approximation$\ssymbol{7}$, direct formulae$\ssymbol{8}$, and Monte Carlo simulations$\ssymbol{9}$.}
	\label{tab:exact-approx-literature}
\end{table}

Historically, applications of goodness-of-fit tests were somehow hampered due to the absence of exact distribution theory for finite sample sizes. Statisticians focused on giving extensive tables of critical values for each statistic's exact distribution and, alternatively, approximating exact distributions of remarkable statistics. Table \ref{tab:exact-approx-literature} lists the approximations available for the exact distributions of $D_n$, $W^2_n$, $V_n$, $U^2_n$, and $A^2_n$, as well as the main techniques behind them. Although these specific approximations are highly accurate, the complexity of their expressions, and the lack of straightforward applicability to other statistics beyond the ones they were designed for, have not displaced the customary use of Monte Carlo simulations, asymptotic distributions, or even lookup tables when emitting general test decisions. In order to reduce the size of lookup tables, \cite{Stephens1970} transformed several statistics $T_n$ (among others, $D_n$, $V_n$, $W_n^2$, and $U_n^2$) into $T_n^{\ast}$ in such a way that the upper tails of $T_n^{\ast}$ remain roughly constant on $n$. Comparing $T_n^{\ast}$ (and not $T_n$) with certain fixed asymptotic critical values for $T_n$ gives a more accurate test calibration for small-to-moderate $n$'s. This approach also allowed finding finite-sample approximations in a wider set of goodness-of-fit problems: \cite{Stephens1974,Stephens1977b,Stephens1979b} and \cite{D'Agostino1986} derived analogous transformations for $D_n$, $V_n$, $W_n^2$, $U_n^2$, and $A_n^2$ when testing the goodness-of-fit of normal, exponential, logistic, and extreme value distributions. Other authors, such as \cite{Dufour1978}, found modifications for $D_n$ to use with truncated or censored samples, and \cite{Crown2000} applied this method to an $A_n^2$-related statistic for testing normal and exponential distributions. \cite{Hegazy1975} found transformations for new test statistics by fitting a functional relationship between the critical values and the sample size, introducing the first explicit use of a regression view to stabilize test statistics and offering insight into Stephens' original work. \cite{Pettitt1977} also applied this regression approach to $A_n^2$ for normality tests. \cite{Johannes1980} proposed an improved modification for \cite{Durbin1969}’s $C$ statistic, finding a specific transformation for each significance level; these approximations give more accurate results for a wider set of significance levels, yet at the expense of tabulating a higher number of transformations. More recently, using several regressions for different significance levels too, \cite{Marks1998, Marks2007} found transformations for $D_n$ to test for Erlang distributions, while \cite{Heo2013} did the same for $A_n^2$ with several extreme value distributions. As Table \ref{table:code-mod} shows, Stephens' transformations are present in nowadays' R software for goodness-of-fit testing, which also implements some of the statistic-specific approaches from Table~\ref{tab:exact-approx-literature}.

\begin{table}[ht!]
	\iffigstabs
	\small
	\centering
	\begin{tabular}{ >{\arraybackslash}p{2.5cm} | >{\centering\arraybackslash}m{1.75cm} | >{\arraybackslash}m{10cm}}
			\toprule
			Methodology & R package & Statistics and references \\
			\midrule
			\multirow{2}{2.2cm}{Exact distributions} & \texttt{goftest} & $W_n^2$ \citep{Csorgo1996}, $A_n^2$ \citep{Marsaglia2004}\\
			\cline{2-3}
			& \texttt{stats} & $D_n$ \citep{Marsaglia2003} \\
			\midrule
			\multirow{3}{2.2cm}{Transformation-based} & \texttt{circular} & $V_n$, $U_n^2$ \citep{Stephens1970}\\
			\cline{2-3}
			& \texttt{sphunif} & $D_n$, $W_n^2$, $V_n$, $U_n^2$ \citep{Stephens1970} \\
			\cline{2-3}
			& \texttt{EnvStats} & $D_n$, $W_n^2$, $A_n^2$ \citep{D'Agostino1986} \\
			\bottomrule
	\end{tabular}
	\fi
	\caption{\small R packages implementing different approximation methods to compute exact $p$-values of goodness-of-fit tests: \texttt{circular} \citep{Agostinelli2017}, \texttt{sphunif} \citep{Garcia-Portugues2020c}, \texttt{EnvStats} \citep{Millard2013}, \texttt{goftest} \citep{Faraway2019}, and \texttt{stats} \citep{RCoreTeam2021}.}
	\label{table:code-mod}
\end{table}

In this paper we build on Stephens' transformations to expand and automate them. First, we present a data-driven procedure to achieve a better stabilization, with respect to the sample size $n$, of the exact null distribution of a generic test statistic $T_n$ of interest, for a wider range of significance levels $\alpha$ (i.e., upper $\alpha$-quantiles of $T_n$). Specifically, new modifications for the (one-sample) Kolmogorov--Smirnov, Cramér--von Mises, Kuiper, and Watson test statistics are derived and shown to extend the scope of applicability of previous approaches. To the best of our knowledge, we also provide the first instance of such a stabilization for the Anderson--Darling test statistic. Second, we provide a method to approximate upper-tail exact $p$-values for the tests constructed from stabilized statistics. Through an extensive simulation study, we show a significant improvement in the precision of the stabilization of the exact critical values of $T_n$ for several sample sizes, as well as a competitive computational cost when compared with statistic-specific methods for evaluating exact null distributions. We also show large improvements, both in precision and computational efficiency, over the use of Monte Carlo simulation, arguably the most popular test calibration approach nowadays. Third, we develop an extension of our stabilization procedure to deal with several recent test statistics for assessing uniformity on $\mathbb{S}^{p-1}$, $p \geq 2$, and which hence have dimension-dependent distributions. In particular, we stabilize the exact null distribution of a novel Anderson--Darling test statistic for circular data. Finally, the introduced stabilization methodology allows us to perform tests in batches of small-to-moderate samples in an accurate and fast manner that does not require Monte Carlo simulation. This is illustrated in an astronomical dataset comprised of the longitudes at which sunspots appear, which exhibits a suspected temporal mix of uniform and non-uniform patterns.

The rest of the paper is organized as follows. Section \ref{section:modification} introduces Stephens' approach (Section \ref{section:Stephens}) and our proposed extension (Section \ref{section:n-alpha-mod}), together with simulation studies and a comparison between several modifications (Section \ref{section:simulations}). Section \ref{section:parameter-modification} briefly introduces the projected-ecdf statistics for testing uniformity on the hypersphere (Section \ref{section:proj-based-unif-tests}), develops the parameter-dependent transformations to achieve their stabilization (Section \ref{section:modification-projected-ecdf}), and analyzes the empirical performance of these transformations (Section \ref{section:simulations2}). Section \ref{section:applications} gives an application of the modified statistics in astronomy. A final discussion of the obtained results concludes the paper in Section \ref{section:discussion}. Further analyses and empirical results are included in the Supplementary Material (SM). All the code and data are available at \url{https://github.com/afernandezdemarcos/approxstats}.

\section{Stabilization of ecdf statistics}
\label{section:modification}

\subsection{On Stephens' stabilization}
\label{section:Stephens}

\cite{Stephens1970} stabilization aims to transform a statistic $T_n$ into $T^{\ast}_n$ through a function of $n$, so that the upper $\alpha$-quantiles of $T^{\ast}_n$ are well approximated by the upper $\alpha$-quantiles of $T_\infty$, the random variable distributed as the asymptotic null distribution of $T_n$, for small-to-moderate sample sizes. The transformation can be interpreted as a two-step stabilization. First, in the \textit{quantile ratios stabilization}, $T_n$ is modified to the statistic $T^{\alpha_0\text{-s}}_n$ so that the ratios of $T^{\alpha_0\text{-s}}_n$'s upper $\alpha$-quantiles with respect to a certain reference upper $\alpha_0$-quantile are roughly constant as a function of $n$. Second, in the \textit{asymptotic stabilization}, $T^{\alpha_0\text{-s}}_n$ is transformed into $T^{\ast}_n$ so that the upper $\alpha$-quantiles of $T^{\ast}_n$ are approximately equal to the asymptotic upper $\alpha$-quantiles for small-to-asymptotic sample sizes. For the sake of brevity, and since we are concerned only with upper-tail tests, henceforth we will use ``$\alpha$-quantile'' as a replacement for ``upper $\alpha$-quantile''.

The ratios involved in the first step are $T_{n; \alpha}/T_{n; \alpha_0}$, where $T_{n; \alpha}$ is the $\alpha$-quantile of the distribution for sample size $n$, i.e., $\mathbb{P}\left[T_{n} \ge T_{n; \alpha}\right] = \alpha$. Obviously, these ratios do not have to be constant for all $n$, as Figure \ref{fig:W2} shows for $W^2_n$. The \textit{quantile ratios stabilization} step searches for a transformed statistic, $T_{n}^{\alpha_0\text{-s}}$, whose quantile ratios $T_{n; \alpha}^{\alpha_0\text{-s}}/T_{n; \alpha_0}^{\alpha_0\text{-s}}$ do not depend on $n$. In other words, the desideratum is that these quantile ratios, for any sample size $n$, equal the asymptotic quantile ratios $T_{\infty; \alpha}/T_{\infty; \alpha_0}$, where $T_{\infty; \alpha}$ is the asymptotic $\alpha$-quantile. One way to find such transformation is by setting $T_{n}^{\alpha_0\text{-s}}:=T_{n} - p(n)$ for a certain function $p:\mathbb{N}\rightarrow \mathbb{R}$ such that it verifies $\lim_{n \to \infty}p(n) = 0$ and the second equality below, for all $n$ and $\alpha$:
\begin{align}
	\frac{T_{n; \alpha}^{\alpha_0\text{-s}}}{T_{n; \alpha_0}^{\alpha_0\text{-s}}} = \frac{T_{n; \alpha} - p(n)}{T_{n; \alpha_0} - p(n)} = \lim_{n \to \infty}\frac{T_{n; \alpha} - p(n)}{T_{n; \alpha_0} - p(n)} 
	= \frac{T_{\infty; \alpha}}{T_{\infty; \alpha_0}} =: k_{\infty; \alpha}. \label{eq:stabilize_ratios}
\end{align}
Hence, $p$ is such that
\begin{align*}
	p(n) = \frac{T_{n; \alpha} - k_{\infty; \alpha} \cdot T_{n; \alpha_0}}{1-k_{\infty; \alpha}},
\end{align*}
which clearly depends on $\alpha$. Stephens fitted $p$ (see the end of this section) for a specific value of $\alpha$, at the expense of accuracy for other significance levels. Upon this step, the quantile ratios of $T_{n}^{\alpha_0\text{-s}}$ are roughly constant for all $n$, as Figure \ref{fig:W2} shows for $W^2_n$. This first step can be omitted for statistics with quantile ratios that are already roughly stable, as it is remarkably the case of $D_n$ and $V_n$ \citep[Section 5]{Stephens1970}. In this case, $p\approx 0$.

The \textit{asymptotic stabilization} step aims to transform the already modified statistic, $T_n^{\alpha_0\text{-s}}$, into $T_n^{\ast}$ so that the $\alpha$-quantiles of this latter statistic are well approximated by the asymptotic $\alpha$-quantiles of the original statistic $T_n$. For that goal, $g:\mathbb{N}\rightarrow \mathbb{R}$ is defined as $g(n):=T_{\infty;\alpha}/T_{n;\alpha}^{\alpha_0\text{-s}}$. Owing to \eqref{eq:stabilize_ratios}, in principle this function does not depend on the significance level $\alpha$, only on $\alpha_0$:
\begin{align}
	\label{eq:asint_ratios_relationship}
	\frac{T_{\infty; \alpha}}{T_{n; \alpha}^{\alpha_0\text{-s}}} = \frac{T_{\infty; \alpha_0}}{T_{n; \alpha_0}^{\alpha_0\text{-s}}},
\end{align}
which holds for any value of $\alpha$. However, when $p$ and $g$ are fitted in practice, \eqref{eq:asint_ratios_relationship} will approximately hold for a certain set of $\alpha$ values, as those shown in Figure \ref{fig:W2}. The function $g$ is estimated from the ratio between $T_{\infty;\alpha}/T_{n;\alpha}^{\alpha_0\text{-s}}$ for a particular value $\alpha_1$ (possibly different from $\alpha_0$), which could result in a loss of accuracy for other quantiles.

The final modified form of $T_n$ is
\begin{align}
	T_n^{\ast} = T^{\alpha_0\text{-s}}_{n} \cdot g(n) = \left(T_n - p(n)\right) \cdot g(n), \label{eqn:stephens-modification}
\end{align}
where we highlight that in practice the functions $p$ and $g$ have to be estimated beforehand. Once these fits are readily available, the main benefit of \eqref{eqn:stephens-modification} is the simplicity of its use, which only requires evaluating a simple $n$-dependent transformation of $T_n$. The fits of $p$ and $g$ were originally handcrafted on a case-by-case basis \citep[even ``found by trial'',][Section 5]{Stephens1970}, or were heavily influenced by Stephens' functional forms, which pose significant limitations in terms of automation and flexibility. Moreover, the approximation error to the exact quantiles of $T_n$ that is obtained is, first, dependent on $\alpha_1$ and, second, significant for $\alpha$-quantiles different from $\alpha_1$. An additional downside of \eqref{eqn:stephens-modification} is the initial stabilization step, which increases the complexity and tuning required (selection of $\alpha_0$), and is a source of uncertainty for the final approximation. In order to overcome these problems, we present in the next section an enhanced stabilization approach that improves the accuracy of exact $\alpha$-quantiles while retaining the simplicity of the transformation.

\begin{figure}[ht!]
	\iffigstabs
	\centering
	\begin{subfigure}{.25\textwidth}
		\centering
		\includegraphics[width=\linewidth]{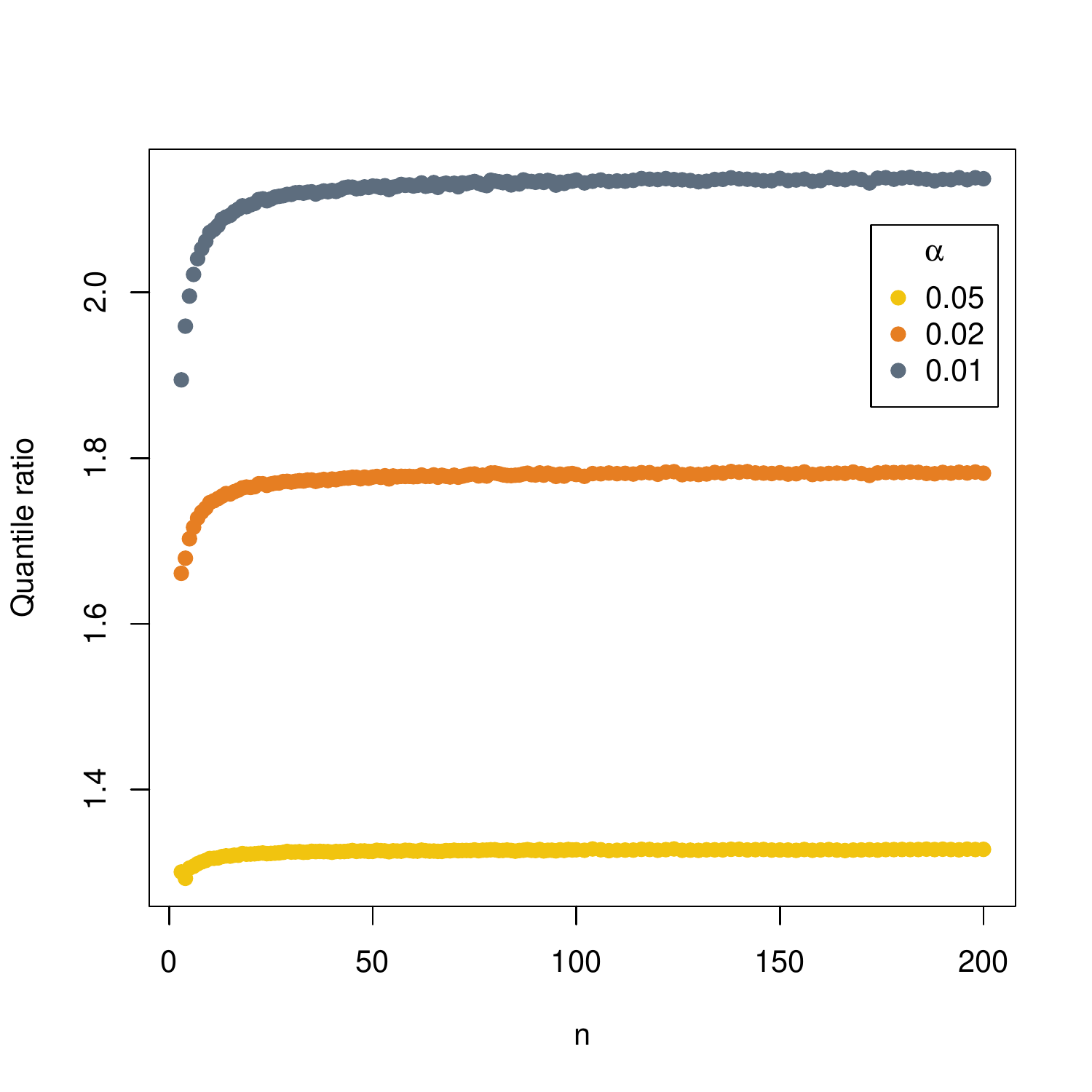}
		\caption{\small $W_{n; \alpha}^2/W_{n; 0.10}^2$}
		\label{fig:W2-ratio}
	\end{subfigure}%
	\begin{subfigure}{.25\textwidth}
		\centering
		\includegraphics[width=\linewidth]{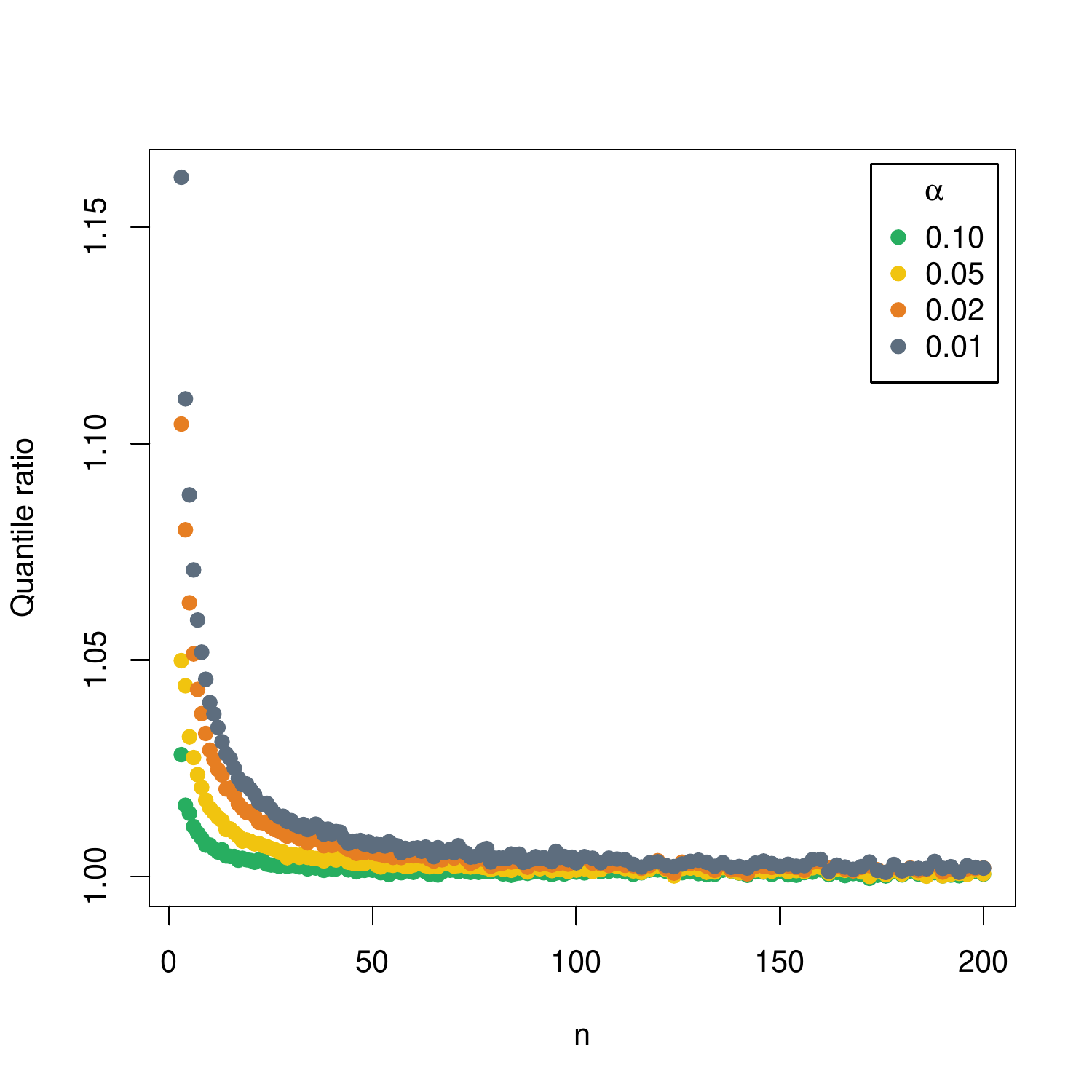}
		\caption{\small $W_{\infty;\alpha}^2/W_{n;\alpha}^2$}
		\label{fig:W2-asint}
	\end{subfigure}%
	\begin{subfigure}{.25\textwidth}
		\centering
		\includegraphics[width=\linewidth]{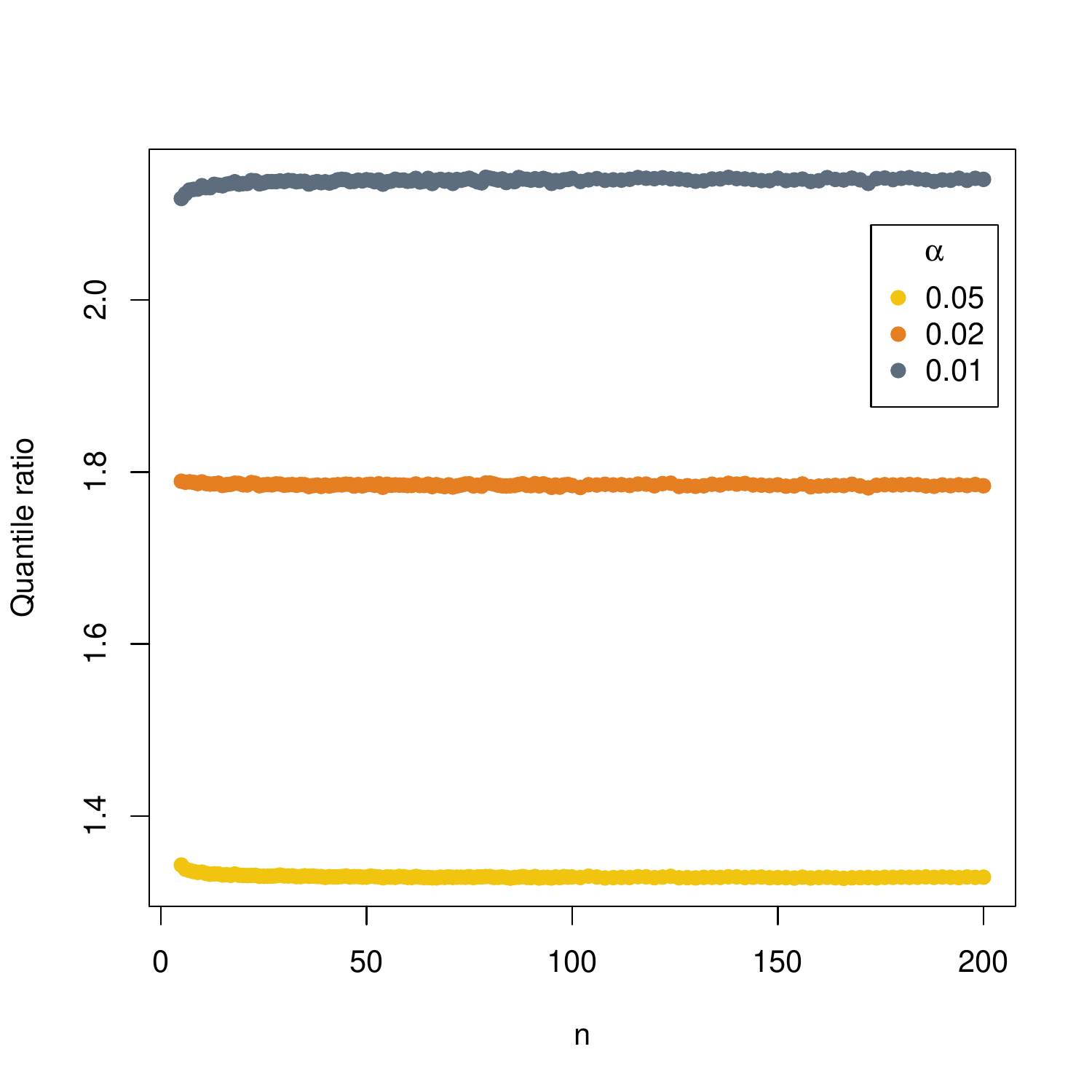}
		\caption{\small $W_{n; \alpha}^{2,0.10\text{-s}}/W_{n; 0.10}^{2,0.10\text{-s}}$}
		\label{fig:W2-stable-ratio}
	\end{subfigure}%
	\begin{subfigure}{.25\textwidth}
		\centering
		\includegraphics[width=\linewidth]{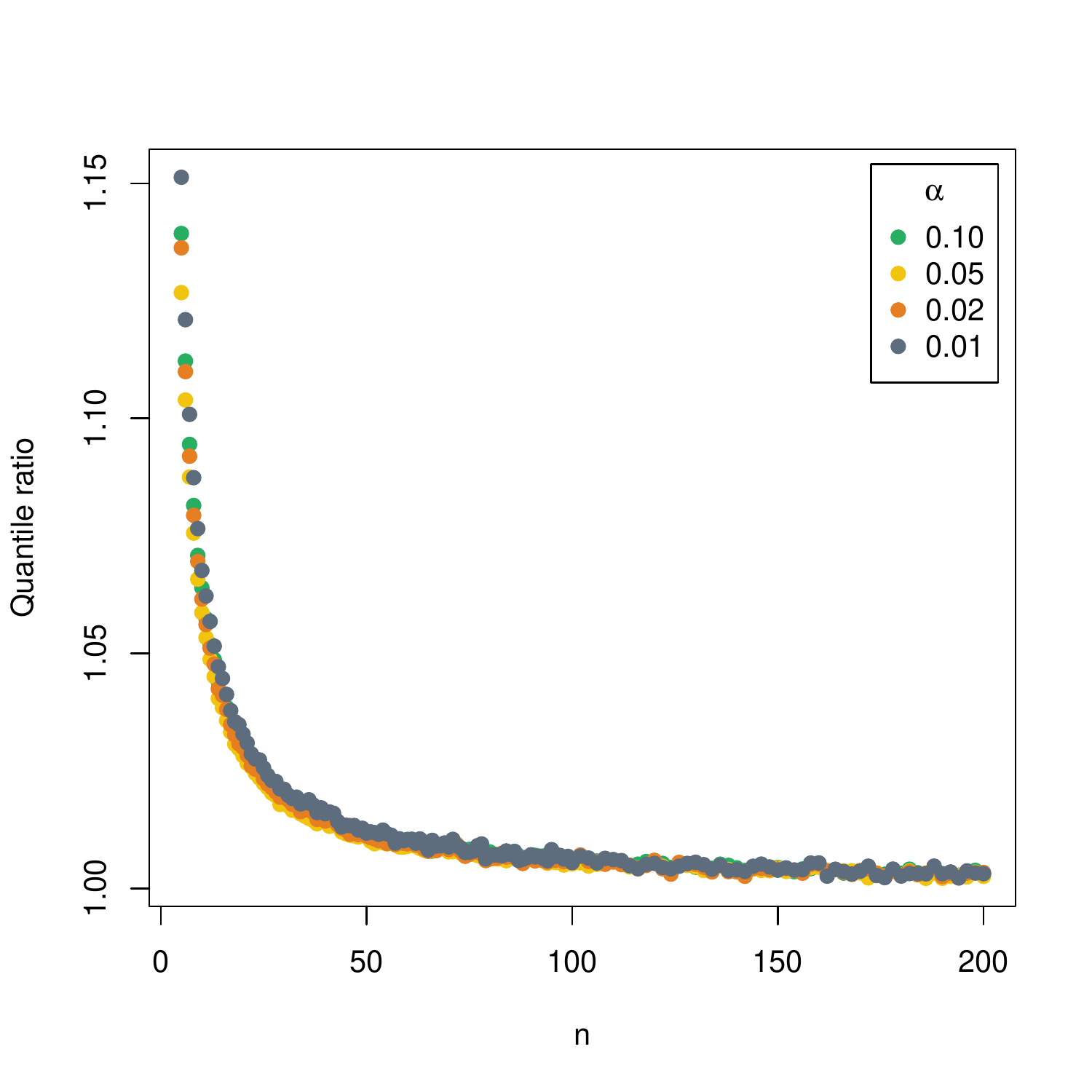}
		\caption{\small $W_{\infty;\alpha}^2/W_{n;\alpha}^{2,0.10\text{-s}}$}
		\label{fig:W2-stable-asint}
	\end{subfigure}
	\fi
	\caption{\small Quantile ratios of the Cramér--von Mises statistic $W_{n}^{2}$ (leftmost two figures) and its ratio-stabilized statistic $W_{n}^{2,0.10\text{-s}}$ (rightmost two figures).}
	\label{fig:W2}
\end{figure}

\subsection{\texorpdfstring{$(n,\alpha)$-stabilization}{(n, alpha)-stabilization}}
\label{section:n-alpha-mod}

Our stabilization consists of a single-step transformation of the original statistic $T_n$ into $T_n^{\ast}(\alpha)$ by a function that depends on the sample size $n$ and the significance level $\alpha$ at which the test is to be performed, so that the exact $\alpha$-quantile of $T_n^{\ast}(\alpha)$ is closely approximated by the $\alpha$-quantile of $T_{\infty}$. Additionally to its improved accuracy and simplicity, an advantage of our modification is that it compresses extensive lookup tables: critical values do not need to be available for different significance levels because $\alpha$ is already included within the transformation.

The ratios $T_{\infty; \alpha}/T_{n; \alpha}$, shown in Figure \ref{fig:W2} for $W_n^2$, can be directly modeled as a function $g:\mathbb{N}\times(0,1)\to\mathbb{R}$ of $(n,\alpha)$, hence condensing the two steps from Section \ref{section:Stephens} into one. To that aim, we define $g$ as the function satisfying
\begin{align}
	\label{eq:f_n_alpha}
	\alpha = \mathbb{P} \left[T_n \ge T_{n; \alpha}\right] = \mathbb{P} \left[T_n\ge T_{\infty; \alpha} / g(n, \alpha)\right],
\end{align}
for all $(n,\alpha)$, and our transformed statistic (for the $\alpha$ significance level) as
\begin{align*}
	T_{n}^{\ast}(\alpha):= T_{n} \cdot g(n, \alpha).
\end{align*}
It is very convenient to reexpress $g$, as defined in \eqref{eq:f_n_alpha}, as
\begin{align}
	\frac{T_{\infty; \alpha}}{T_{n; \alpha}} = g(n, \alpha) + \varepsilon,\label{eq:reg}
\end{align}
where $\varepsilon=0$ if \eqref{eq:f_n_alpha} is perfectly satisfied for all $(n,\alpha)$. Indeed, Equation \eqref{eq:reg} casts the stabilization problem as an error-free fixed-design regression problem with predictors $(n,\alpha)$, response $Y:=T_{\infty; \alpha}/T_{n; \alpha}$, and unknown regression function $g$. Casting Stephens' stabilizations as a regression problem was early introduced in \cite{Hegazy1975}, \cite{Pettitt1977}, and \cite{Johannes1980}. Yet these works focus on using the sample size as the unique predictor, for isolated $\alpha$-quantiles, an approach that has been later applied in \cite{Marks1998, Marks2007} and \cite{Heo2013}.

We introduce now a sufficiently flexible parametric specification for $g$ in \eqref{eq:reg} that allows its effective estimation in practice. We resort to a linear model featuring negative powers of the sample size $n$ and significance level $\alpha$ as predictors, as well as the corresponding interaction effects between them. Precisely, we consider the following saturated model:
\begin{align}
	g(n, \alpha)= 1+\frac{\beta_1}{\sqrt{n}}+\frac{\beta_2}{n}
	+\frac{\beta_3}{\sqrt{n\alpha}}+\frac{\beta_4}{\sqrt{n}\alpha}+\frac{\beta_5}{n\sqrt{\alpha}}+\frac{\beta_6}{n\alpha}. \label{eq:saturated_g_n_alpha}
\end{align}
The fixed intercept and negative powers of $n$ were included to guarantee that $\lim_{n\to\infty}g(n,\alpha)=1$, thus in accordance with $\lim_{n\to\infty} T_{\infty; \alpha}/T_{n; \alpha}=1$. Powers of $n^{-1/2}$ resemble the sample size factors in the terms of an Edgeworth series. The powers of $\alpha^{-l/2}$, $l=1,2$, were experimentally found to be an appropriate specification for capturing the interactions with $n$. The appropriateness of the model specification \eqref{eq:saturated_g_n_alpha} is exhaustively investigated in \myref{supplementary:powers}{Appendix A} in the SM. Upon available samples of the form $\{(n_j,\alpha_j,Y_j)\}_{j=1}^J$, $Y_j:=T_{\infty; \alpha_j}/T_{n_j; \alpha_j}$, model \eqref{eq:saturated_g_n_alpha} is estimated through weighted least squares, using the weight $\smash{w_j:=n_j^{-1/2}1_{\{0<\alpha_j\leq 0.25\}}}$ for the $j$-th observation to give heavier weight to the approximation error on lower sample sizes. The indicator in $w_j$ reflects our interest in only stabilizing the upper tail of the test statistic $T_n$, hence disregarding those quantiles associated with non-rejections of the test based on $T_n$. \myref{supplementary:weights}{Appendix B} in the SM provides more detail on the selection of the weight function among several alternatives.

The data required for fitting \eqref{eq:saturated_g_n_alpha} is to be produced under the (fairly realistic nowadays) assumption that it is feasible to simulate a large number of statistics $T_n$ under the null hypothesis and for varying sample sizes. Specifically, we have carried out the following simulation for the test statistics $D_n$, $W^2_n$, $V_n$, $U_n^2$, and $A_n^2$. We produced $M=10^7$ Monte Carlo random samples for $T_n$, for each of the sample sizes $n$ in the set $\mathcal{N}:=\{5,\ldots,100, 102,\ldots, 200, 204, \ldots,300,308,\ldots,404,\allowbreak420,\allowbreak\ldots,500\}$. We then condensed these statistics as the quantiles $\{T_{n_j;\alpha_j}: n_j\in\mathcal{N},\,\alpha_j\in\mathcal{A}\}$, for $\mathcal{A}:=\{a/A:a=1,\ldots,A\}$, $A=10^3$. The asymptotic $\alpha$-quantiles $\{T_{\infty; \alpha_j}:\alpha_j\in\mathcal{A}\}$ were computed from the statistics' asymptotic null distributions, as those were readily available in the literature. The generated sample is therefore $\{(n_j,\alpha_j,Y_j)\}_{j=1}^J$, $J=\#\mathcal{N}\times A$. Clearly, this is a computationally-intensive process, although it only needs to be done once per kind of test statistic. The procedure is analogous for other one-sample test statistics that are feasible to simulate under the simple null hypothesis at hand. If the limiting distribution is not available or tractable, a sufficiently large sample size $n$ could be used to approximate $T_{\infty; \alpha}$ by $T_{n; \alpha}$ by Monte Carlo.

Using the sample $\{(n_j,\alpha_j,Y_j)\}_{j=1}^J$, we advocate the use of stepwise regression for performing model selection within \eqref{eq:saturated_g_n_alpha}. Specifically, we performed a forward-backward search for minimizing the Bayesian Information Criterion (BIC) on the space of models contained in \eqref{eq:saturated_g_n_alpha}. The search was initiated with the model featuring only the predictors used in Stephens' modifications, i.e., $n^{-1/2}$ and $n^{-1}$. To attain simpler models than the BIC-optimal one, a final step was implemented to iteratively drop one-by-one the predictors that contributed the least to the adjusted $R^2$ of the resulting model. The threshold was established to keep only three final terms (for simplicity), the predictors removed decreasing less than $0.15\%$ the $R^2_\mathrm{adj}$ which, averaged across the five statistics, was larger than $0.96$.

\begin{table}[ht!]
	\iffigstabs
	\centering
	\small
	\begin{tabular}{>{\centering\arraybackslash}m{0.5cm} >{\centering\arraybackslash}m{5cm} >{\centering\arraybackslash}m{1cm} >{\centering\arraybackslash}m{1cm} >{\centering\arraybackslash}m{1cm} >{\centering\arraybackslash}m{1cm} >{\centering\arraybackslash}m{1cm}} 
		\toprule
		$T_n$ & $T_n^{\ast}(\alpha)$ & $T_{\infty;0.15}$ & $T_{\infty;0.1}$ & $T_{\infty;0.05}$ & $T_{\infty;0.025}$ & $T_{\infty;0.01}$\\ 
		\midrule
		$D_n$ & $D_n\left(1 + \frac{0.1575}{\sqrt{n}} + \frac{0.0192}{n\sqrt{\alpha}} - \frac{0.0051}{\sqrt{n\alpha}}\right)$ & $1.1380$ & $1.2239$ & $1.3581$ & $1.4803$ & $1.6277$ \\ [3ex]
		$W_n^2$ & $W_n^2\left(1 - \frac{0.1651}{n} + \frac{0.0749}{n\sqrt{\alpha}} - \frac{0.0014}{n \alpha}\right)$ & $0.2841$ & $0.3473$ & $0.4613$ & $0.5806$ & $0.7435$ \\ [3ex]
		$V_n$ & $V_n\left(1 + \frac{0.2330}{\sqrt{n}} + \frac{0.0276}{n\sqrt{\alpha}} - \frac{0.0068}{\sqrt{n\alpha}}\right)$ & $1.5370$ & $1.6196$ & $1.7473$ & $1.8625$ & $2.0010$ \\ [3ex]
		$U_n^2$ & $U_n^2\left(1 - \frac{0.1505}{n} + \frac{0.0917}{n\sqrt{\alpha}} - \frac{0.0018}{n \alpha}\right)$ & $0.1313$ & $0.1518$ & $0.1869$ & $0.2220$ & $0.2685$ \\ [3ex]
		$A_n^2$ & $A_n^2\left(1 + \frac{0.0360}{n} - \frac{0.0234}{n \sqrt{\alpha}} + \frac{0.0006}{n \alpha}\right)$ & $1.6212$ & $1.9331$ & $2.4922$ & $3.0775$ & $3.8784$ \\ [3ex]
		\bottomrule
	\end{tabular}
	\fi
	\caption{\small Modified statistics for sample size $n$ and significance level $\alpha$. Modified forms are valid for $n \ge 5$ and $0<\alpha \leq 0.25$. $\mathcal{H}_0$ is rejected at significance level $\alpha$ if $T_{n}^{\ast}(\alpha)>T_{\infty;\alpha}$.}\label{table:modified-stats}
\end{table}

The resulting modified forms for $D_n$, $W_n^2$, $V_n$, $U_n^2$, and $A_n^2$ are collected in Table \ref{table:modified-stats}. All of the transformations have three correcting terms, one dependent on $n$ and the other two related to $n$ and $\alpha$, $(n\sqrt{\alpha})^{-1}$ being common to the five statistics. Interestingly, the same correction terms are present within the groups of supremum- and quadratic-norm statistics, as well as in the pairs of linear and circular variants. These forms are valid for $n \ge 5$, which anecdotally gives a minor improvement over Stephens' forms, valid for $n\ge 8$. The steps to use them with the upper-tail test for $\mathcal{H}_0$ that is based on $T_n$ and that is carried out at the significance level $\alpha$ are as follows:
\begin{enumerate}[label=(\textit{\roman*}),ref=(\textit{\roman*})]
	\item Compute the test statistic $T_n$ using its original form.
	\item Calculate the corresponding modified test statistic, $T_n^{\ast}(\alpha)$, in Table \ref{table:modified-stats}.
	\item Retrieve an asymptotic critical value $T_{\infty;\alpha}$ in Table \ref{table:modified-stats}. If $T_n^{\ast}(\alpha)>T_{\infty;\alpha}$, reject $\mathcal{H}_0$ at significance level $\alpha$.
\end{enumerate}

The transformed statistics can also be used to obtain approximations to exact $p$-values, provided the asymptotic quantiles $\mathcal{T}_\infty:=\{T_{\infty;\alpha_j}:\alpha_j\in\mathcal{A}\}$ have been precomputed. This is done in two steps. First, $p$-value bounds $[\alpha_1,\alpha_2]$ are obtained from the grid $\mathcal{A}$ such that $T_n^*(\alpha_1)\leq T_{\infty;\alpha_1}$ and $T_n^*(\alpha_2)> T_{\infty;\alpha_2}$. Once these discrete bounds for $p\text{-value}$ are available, a linear interpolation is applied to define $t_\infty(\alpha):=T_{\infty;\alpha_1}+(T_{\infty;\alpha_2}-T_{\infty;\alpha_1})(\alpha-\alpha_1)/(\alpha_2-\alpha_1)$ for $\alpha\in[\alpha_1,\alpha_2]$ and then the root $\alpha^*\in[\alpha_1,\alpha_2]$ of
\begin{align}
	T_n^*(\alpha^*)=t_\infty(\alpha^*)\label{eq:root}
\end{align}
is obtained by Newton--Raphson (NR). The approximate $p\text{-value}$ is then set to $\alpha^*$. If $\alpha_1 \ge \alpha_{\max}$, $\alpha_{\max}=0.25$ being the maximum element in $\mathcal{A}$ for which the transformation has been estimated, $p\text{-value}=\alpha_{\max}$ is returned. Algorithm \ref{alg:p-val} summarizes this process.

\begin{algorithm}
	\caption{\small $p$-value approximation using the $(n, \alpha)$-modification}\label{euclid}
	\label{alg:p-val}
	\small
	\begin{algorithmic}[1]
		\Function{pvalue\_approx}{$T_n$, $n$, $\mathcal{T}_\infty$, $\mathcal{A}$}
		\For{$j \textbf{ from } 1\textbf{ to }\#\mathcal{A}$}
		\State $T_{\mathrm{mod}, \alpha} \gets T_n^*(T_n, n, \mathcal{A}\left[j\right])$
		\If {$T_{\mathrm{mod}, \alpha} > \mathcal{T}_\infty\left[j\right]$}
		\If {$j=1$}
		\State $\left(\alpha_1, \alpha_2\right) \gets \left(\mathcal{A}\left[j\right], \mathcal{A}\left[j+1\right]\right)$
		\State $\left(T_{\infty;\alpha_2},T_{\infty;\alpha_1}\right) \gets \left(\mathcal{T}_\infty\left[j\right], \mathcal{T}_\infty\left[j+1\right]\right)$
		\Else
		\State $\left(\alpha_1, \alpha_2\right) \gets \left(\mathcal{A}\left[j-1\right], \mathcal{A}\left[j\right]\right)$
		\State $\left(T_{\infty;\alpha_2},T_{\infty;\alpha_1}\right) \gets \left(\mathcal{T}_\infty\left[j-1\right], \mathcal{T}_\infty\left[j\right]\right)$
		\EndIf
		\State $\alpha^\ast \gets \operatorname{NR}(T_n^*(T_n, n, \alpha)-t_\infty(\alpha, \mathcal{T}_{\infty}, \alpha_1, \alpha_2))$
		\State $\textbf{return }\alpha^\ast$
		\EndIf
		\EndFor
		\State $\textbf{return } 0.25$
		\EndFunction
	\end{algorithmic}
\end{algorithm}

When there is no $\alpha_1$ in $\mathcal{A}$ such that $T_n^*(\alpha_1)\leq T_{\infty;\alpha_1}$, the $p$-value is set as the nonnegative extrapolation of the root in \eqref{eq:root}, with $\alpha_1$ and $\alpha_2$ being the two lowest elements in $\mathcal{A}$.

\subsection{Simulation study}
\label{section:simulations}

\begin{figure}[htpb!]
	\iffigstabs
	\centering
	\vspace{-0.25cm}
	\vspace{-3.25em}
	\begin{subfigure}{0.47\textwidth}
		\includegraphics[width=\linewidth]{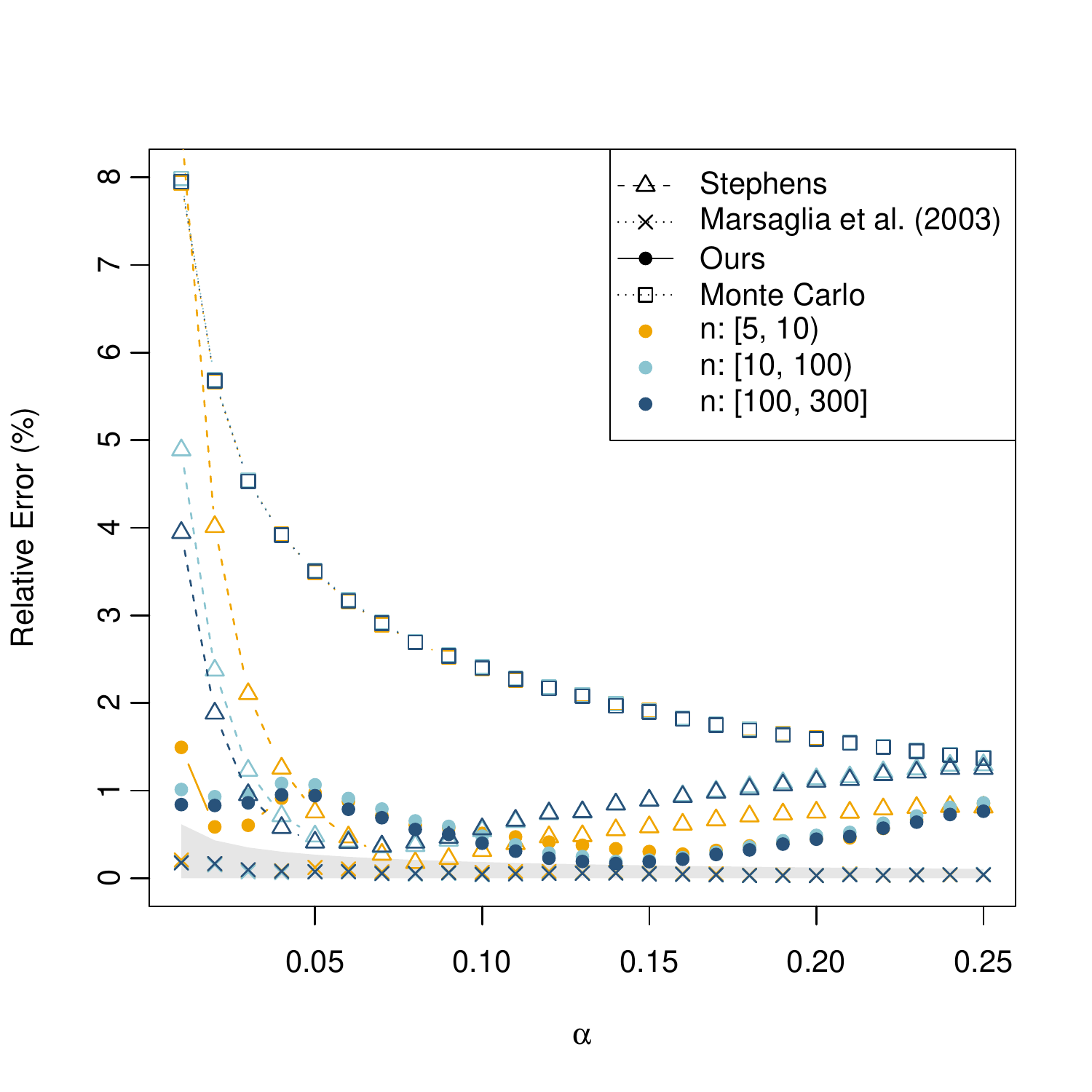}
		\caption{\small $D_n$\label{fig:Dn}}
	\end{subfigure}%
	\begin{subfigure}{0.47\textwidth}
		\includegraphics[width=\linewidth]{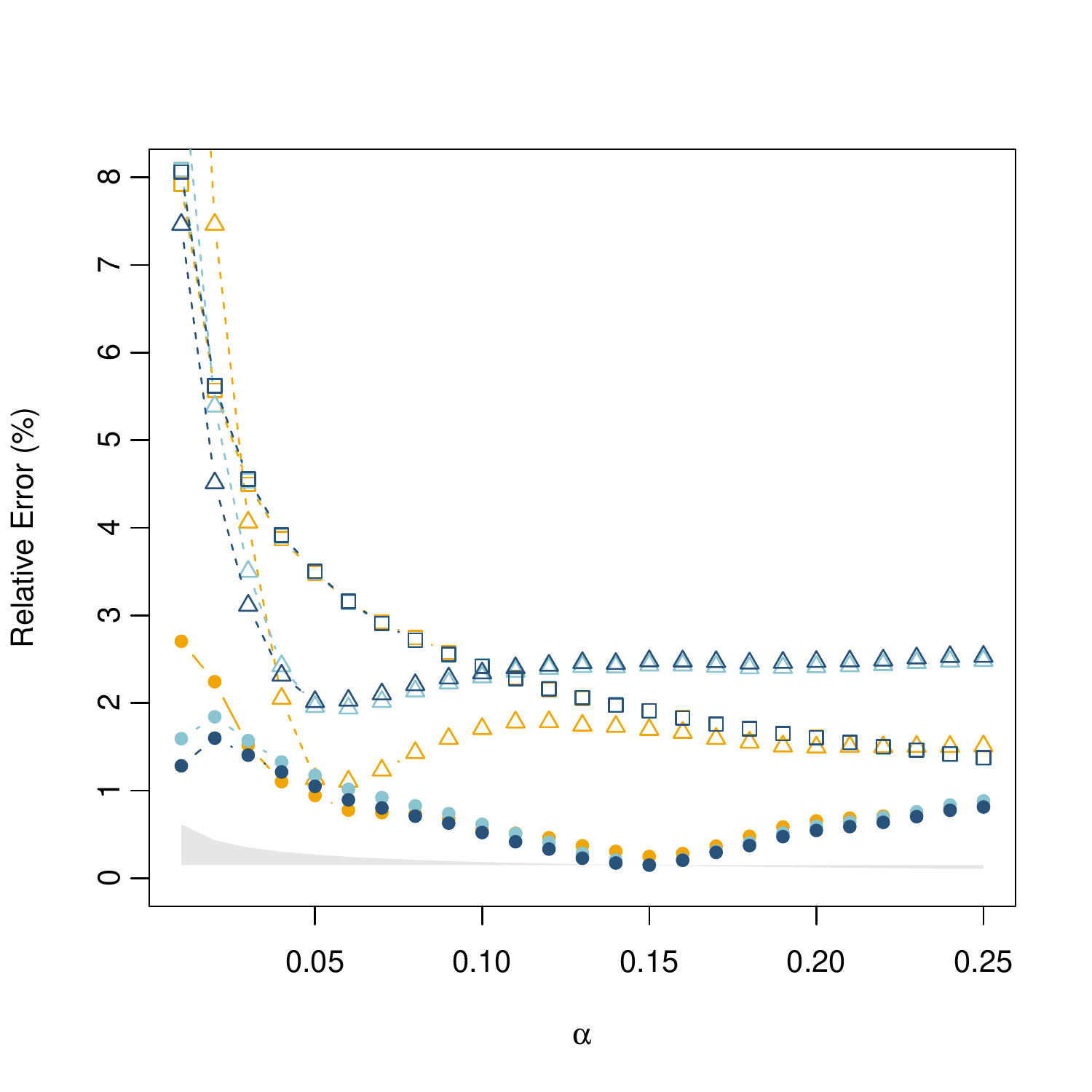}
		\caption{\small $V_n$}
	\end{subfigure}
	\vspace{-0.75cm}

	\begin{subfigure}{0.47\textwidth}
		\includegraphics[width=\linewidth]{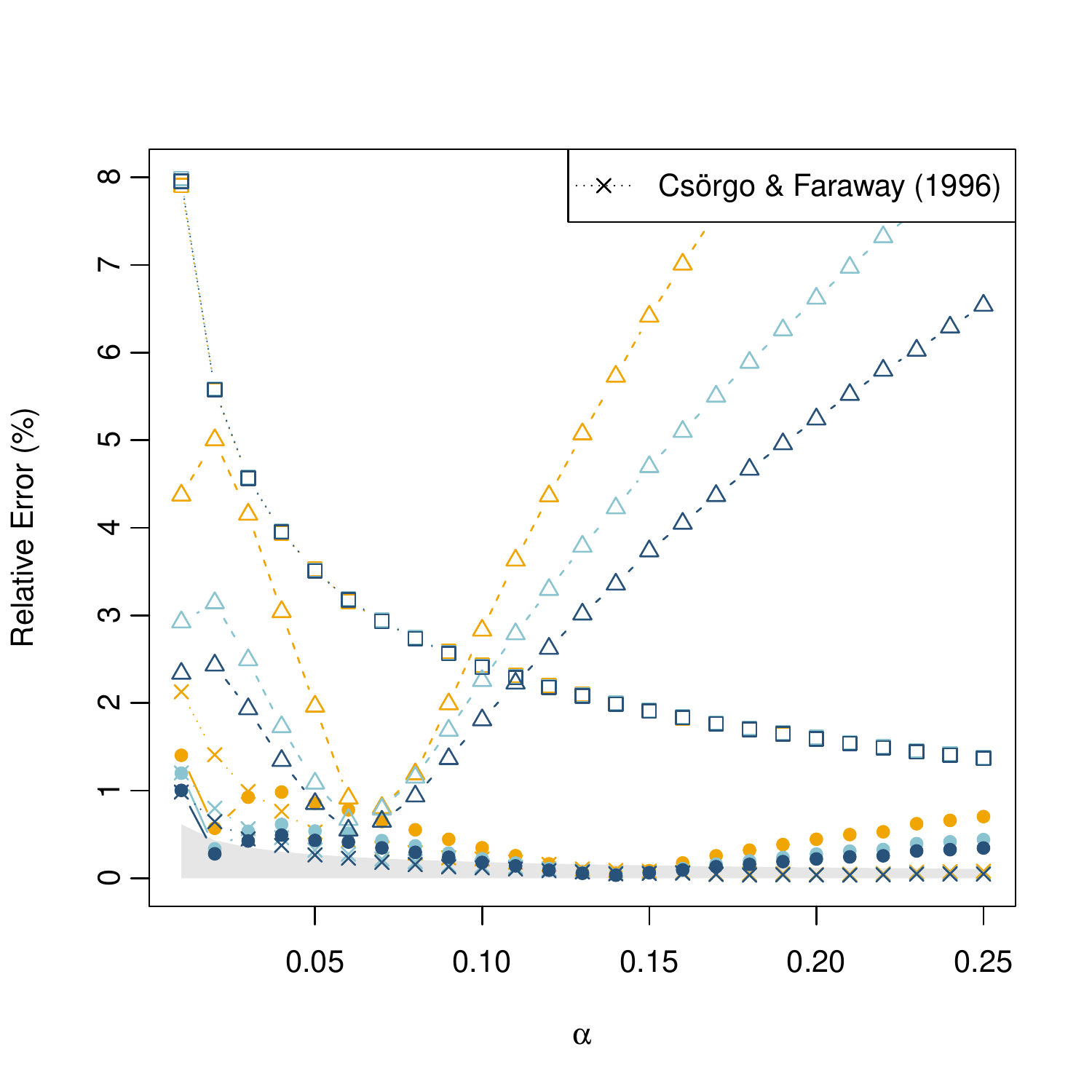}
		\caption{\small $W^2_n$\label{fig:Wn2}}
	\end{subfigure}%
	\begin{subfigure}{0.47\textwidth}
		\includegraphics[width=\linewidth]{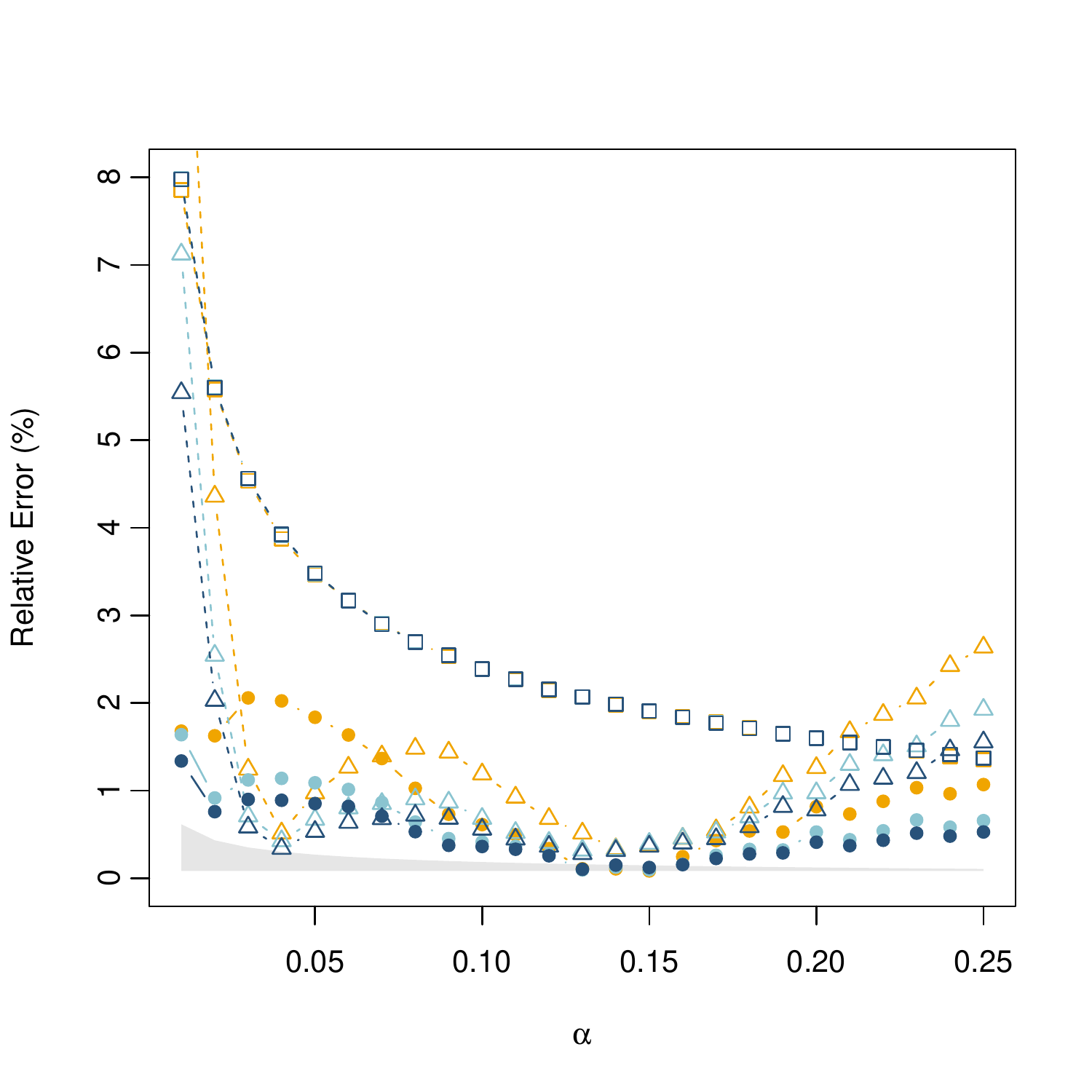}
		\caption{\small $U^2_n$}
	\end{subfigure}
	\vspace{-0.75cm}
	
	\begin{subfigure}{0.47\textwidth}
		\includegraphics[width=\linewidth]{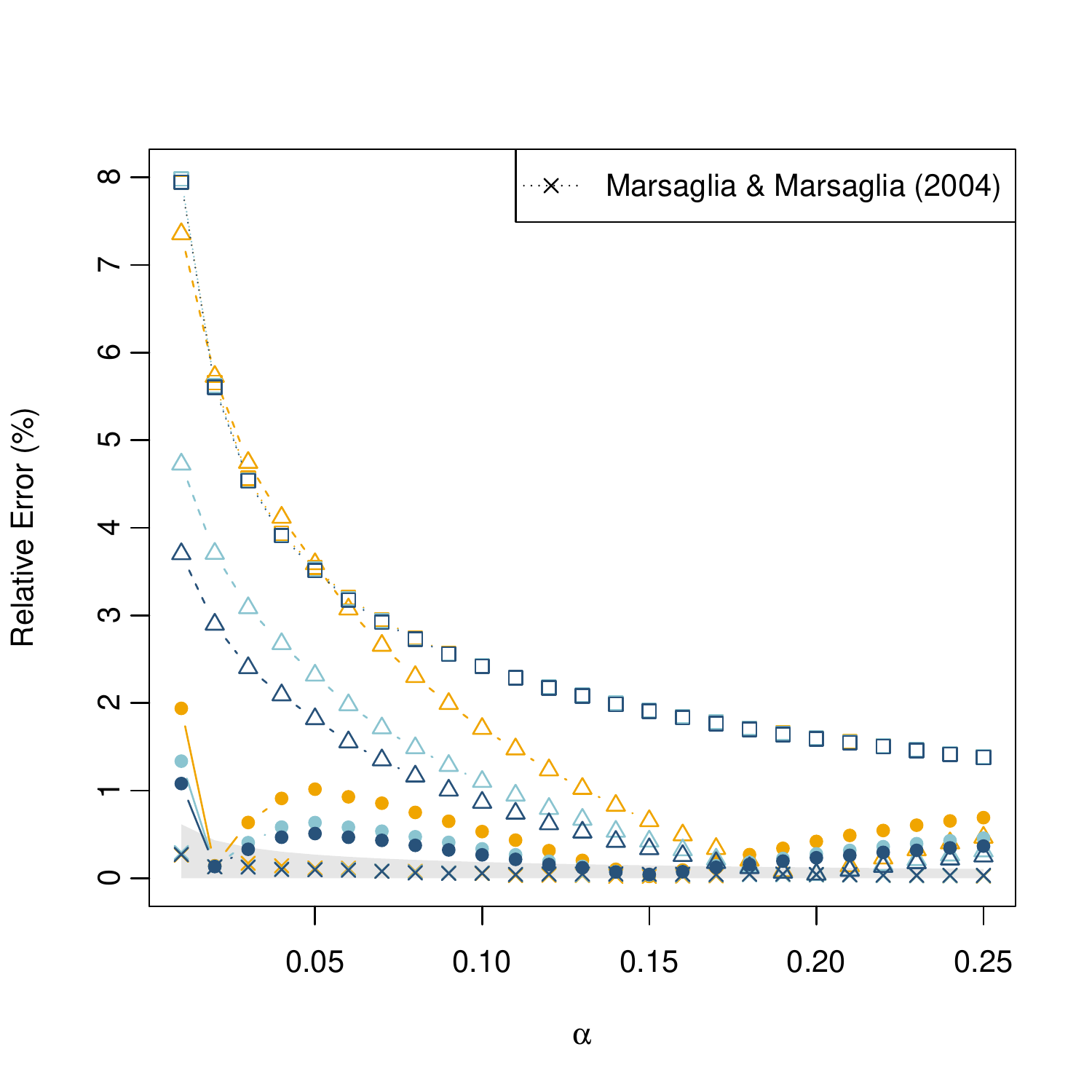}
		\caption{\small $A^2_n$\label{fig:An2}}
	\end{subfigure}
	\fi
	\caption{\small Relative error (in \%) $\lvert\alpha-\tilde{\alpha}\rvert/\alpha$ between the significance level $\alpha$ and $\tilde{\alpha}$, the empirical rejection rate using an approximated exact-$n$ critical value, averaged across different sample sizes $n$. The legend in Figure \ref{fig:Dn} details the approximation methods considered and applies to the rest of the panels, with different specific methods in Figures \ref{fig:Wn2} and \ref{fig:An2}. The gray shaded area corresponds to the $95\%$ confidence interval of the relative error when $\tilde{\alpha}$ is produced by the exact-$n$ critical value estimated by $M = 10^7$ Monte Carlo samples.}
	\label{fig:rejection-proportion-MC-error}
\end{figure}

For the test statistics $D_n$, $V_n$, $W^2_n$, $U^2_n$, and $A_n^2$, we evaluate next the divergence of the exact-$n$ critical values under $\mathcal{H}_0$ from their corresponding approximations given by: (a) Stephens' modified forms; (b) the particular approximation methods from Table \ref{table:code-mod}; (c) Monte Carlo approximation with $10^4$ trials; and (d) our proposed transformations. Figure \ref{fig:rejection-proportion-MC-error} displays the relative errors for the rejection proportions generated by approximated critical values based on methods (a)--(d). These relative errors are defined as $\lvert\alpha-\tilde{\alpha}\rvert/\alpha$, where $\alpha$ is the significance level and $\tilde{\alpha}$ is the empirical rejection rate obtained with $M=10^7$ Monte Carlo samples when using an $\alpha$-critical value computed by each approximation method. The $M=10^7$ Monte Carlo samples under $\mathcal{H}_0$ were drawn for each of the sample sizes $n$ in $\mathcal{N}_{\text{test}}:=\{5, \ldots, 10, 20, \ldots, 50, 100, 200, 300\}$. The sample quantiles for the significance levels in $\mathcal{A}_{\text{test}}:= \{a/100:a=1, \ldots, 25\}$ were computed for each sample size and statistic. For the critical value approximations (a) and (d), critical values were computed by applying the corresponding inverse transformation from Table \ref{table:summary-table} to the asymptotic $\alpha$-critical value $T_{\infty;\alpha}$. Obtaining the critical values in (b) is straightforward using the functions \texttt{stats:::C\_pKolmogorov2x} \citep{RCoreTeam2021} for $D_n$, and \texttt{goftest::pCvM} and \texttt{goftest::pAD} \citep{Faraway2019} for $W^2_n$ and $A^2_n$, respectively. For the critical value approximation based on (c), the (random) relative error for each critical value was averaged over $10^3$ simulations to give an estimate of the average Monte Carlo relative error. Each panel in Figure \ref{fig:rejection-proportion-MC-error} shows the relative error along $\mathcal{A}_{\text{test}}$ averaged for three sets of sample sizes: $5 \leq n < 10$, $10 \leq n < 100$, and $n \geq 100$.

Along $\mathcal{A}_{\text{test}}$, the average relative errors of our stabilizations are $0.5\%$, $0.3\%$, $0.5\%$, $0.3\%$, and $0.7\%$ for $D_n$, $W_n^2$, $U_n^2$, $A_n^2$, and $V_n$, respectively. The relative errors remain fairly stable for every significance value in $\mathcal{A}_{\text{test}}$ without significant differences between the sets of sample sizes analyzed. Compared to Stephens' stabilizations, our relative error is lower by a factor of $\times2$, $\times12$, $\times2$, $\times3$, and $\times4$ on average, respectively. The largest improvements are achieved for $\alpha \neq 0.05$, since Stephens' stabilizations were tuned for $\alpha=0.05$, and for sample sizes $n \leq 100$. This behavior is more obvious in $W_n^2$ and $U_n^2$, which are the statistics that, in Stephens' approach, use an additional prior step for stabilizing the quantile ratios. When compared to the Monte Carlo approximation with $10^4$ samples, our relative error is lower for every significance level and sample size tested, and improves by $\times5$, $\times10$, $\times5$, $\times9$, and $\times4$ on average, respectively. As expected, the approximation methods that are specifically designed for each test statistic achieve the lowest relative errors.

\begin{table}[ht!]
	\iffigstabs
	\small
	\centering
	\begin{tabular}{>{\centering\arraybackslash}m{1cm} wr{0.6cm} wr{0.6cm} wr{0.6cm} wr{0.6cm} wr{0.6cm} wr{0.6cm} wr{0.6cm} wr{0.6cm} wr{0.6cm} wr{0.6cm}} 
		\toprule
		\multirow{2}{*}{$\alpha$} & \multicolumn{10}{c}{$n$}\\ [1ex]
		& \multicolumn{1}{c}{$5$} & \multicolumn{1}{c}{$6$} & \multicolumn{1}{c}{$7$} & \multicolumn{1}{c}{$8$} & \multicolumn{1}{c}{$9$} & \multicolumn{1}{c}{$10$} & \multicolumn{1}{c}{$20$} & \multicolumn{1}{c}{$30$} & \multicolumn{1}{c}{$40$} & \multicolumn{1}{c}{$50$}\\
		\midrule
		\multicolumn{11}{c}{$D_n$: \cite{Marsaglia2003} vs. Algorithm \ref{alg:p-val}} \\ [1ex]
		$0.01$ & $2.48$ & $2.56$ & $3.04$ & $2.96$ & $3.00$ & $3.23$ & $7.09$ & $10.74$ & $17.08$ & $23.38$ \\ 
		$0.02$ & $2.28$ & $2.28$ & $2.40$ & $2.75$ & $2.80$ & $2.85$ & $4.80$ & $9.62$ & $12.04$ & $17.39$ \\ 
		$0.05$ & $1.61$ & $1.97$ & $1.90$ & $1.87$ & $1.90$ & $2.29$ & $3.06$ & $5.94$ & $7.29$ & $10.50$ \\ 
		$0.10$ & $1.22$ & $1.21$ & $1.44$ & $1.43$ & $1.48$ & $1.49$ & $2.24$ & $3.33$ & $4.17$ & $6.05$ \\ 
		$0.15$ & $1.02$ & $0.96$ & $0.98$ & $1.13$ & $1.15$ & $1.19$ & $1.42$ & $2.74$ & $3.45$ & $3.67$ \\ 
		$0.25$ & $0.68$ & $0.71$ & $0.70$ & $0.71$ & $0.70$ & $0.81$ & $1.04$ & $1.48$ & $1.82$ & $2.64$ \\ 
		\midrule
		\multicolumn{11}{c}{$W^2_n$: \cite{Csorgo1996} vs. Algorithm \ref{alg:p-val}} \\ [1ex]
		$0.01$ & $10.43$ & $10.40$ & $10.17$ & $10.12$ & $10.03$ & $10.00$ & $10.60$ & $10.68$ & $10.66$ & $11.82$ \\ 
		$0.02$ & $8.69$ & $8.47$ & $8.42$ & $8.47$ & $8.73$ & $8.75$ & $8.92$ & $9.06$ & $8.85$ & $8.99$ \\ 
		$0.05$ & $5.54$ & $5.53$ & $5.61$ & $5.57$ & $5.56$ & $5.58$ & $5.67$ & $5.68$ & $5.64$ & $5.68$ \\ 
		$0.10$ & $3.46$ & $3.50$ & $3.48$ & $3.46$ & $3.45$ & $3.48$ & $3.50$ & $3.48$ & $3.48$ & $3.49$ \\ 
		$0.15$ & $2.50$ & $2.48$ & $2.49$ & $2.54$ & $2.48$ & $2.55$ & $2.57$ & $2.50$ & $2.51$ & $2.52$ \\ 
		$0.25$ & $1.62$ & $1.62$ & $1.63$ & $1.59$ & $1.59$ & $1.64$ & $1.61$ & $1.61$ & $1.65$ & $1.64$ \\ 
		\midrule
		\multicolumn{11}{c}{$A^2_n$: \cite{Marsaglia2004} vs. Algorithm \ref{alg:p-val}} \\ [1ex]
		$0.01$ & $6.66$ & $6.28$ & $6.23$ & $6.14$ & $6.20$ & $6.42$ & $6.29$ & $6.18$ & $6.29$ & $6.43$ \\ 
		$0.02$ & $6.00$ & $6.52$ & $5.91$ & $6.18$ & $6.13$ & $6.22$ & $5.91$ & $6.26$ & $6.14$ & $6.60$ \\ 
		$0.05$ & $5.12$ & $5.24$ & $5.72$ & $5.74$ & $5.36$ & $6.04$ & $5.24$ & $5.23$ & $5.44$ & $5.44$ \\ 
		$0.10$ & $4.26$ & $4.39$ & $4.35$ & $4.26$ & $4.26$ & $4.81$ & $4.35$ & $4.52$ & $4.36$ & $4.32$ \\ 
		$0.15$ & $3.70$ & $3.62$ & $3.64$ & $3.78$ & $3.64$ & $3.65$ & $3.78$ & $3.75$ & $3.72$ & $3.74$ \\ 
		$0.25$ & $2.87$ & $3.19$ & $2.79$ & $2.85$ & $3.10$ & $3.02$ & $2.83$ & $2.85$ & $2.98$ & $2.87$ \\
		\bottomrule
	\end{tabular}
	\fi
	\caption{\small Running time ratios between specific $p$-value approximation methods and our $p$-value approximation method (Algorithm \ref{alg:p-val}). Ratios are computed for the median running times of $10^3$ evaluations, for each pair $(n, \alpha)$. The averages of the median running times of Algorithm \ref{alg:p-val} are $3.65 \mu s$, $225 \mu s$ (for R version, $4.5 \mu s$ for C++ version), and $3 \mu s$ for $D_n$, $W^2_n$, and $A^2_n$, respectively.}
	\label{table:exec-time}
\end{table}

Table \ref{table:exec-time} presents a comparison of the running times between our $p$-value approximation (Algorithm \ref{alg:p-val}) and the already implemented $p$-value approximation methods for $D_n$, $W^2_n$, and $A^2_n$ described in Table \ref{table:code-mod}. Our method is shown to be $\times 3.8$, $\times 5.4$, and $\times 4.8$ faster than \cite{Marsaglia2003}, \cite{Csorgo1996}, and \cite{Marsaglia2004}, respectively. These methods are already implemented in C++, except for \cite{Csorgo1996} which is in R. Hence, C++ and R versions implementing Algorithm \ref{alg:p-val} were developed for each statistic to allow a fair comparison. In addition, Table \ref{table:exec-timeMC} compares the running times between the $p$-value approximation based on Algorithm \ref{alg:p-val} and a Monte Carlo $p$-value approximation based on $10^4$ trials, which shows that our method is $\times 75 \cdot 10^4$, $\times 58 \cdot 10^4$, and $\times 93 \cdot 10^4$ faster. Monte Carlo approximation was implemented in R code with calls to C++-coded statistics (the most time-consuming part), and the C++ version of Algorithm \ref{alg:p-val} was used. All comparisons were carried out using \texttt{microbenchmark} package \citep{Mersmann2019}. In order to compute the median running time of each function for a given sample size $n$ and significance level $\alpha$, $10^3$ evaluations of the compiled functions were run after $10$ warm-up runs using the same machine, a regular desktop computer with a $3.6$GHz processor. In all cases, the computation of the original statistic $T_n$ was excluded from the timings. R and C++ integration was done with the \texttt{Rcpp} package \citep{Eddelbuettel2011}.

\begin{table}[ht!]
	\iffigstabs
	\small
	\centering
		\begin{tabular}{>{\centering\arraybackslash}m{1cm} >{\centering\arraybackslash}m{0.6cm} >{\centering\arraybackslash}m{0.6cm} >{\centering\arraybackslash}m{0.6cm} >{\centering\arraybackslash}m{0.6cm} >{\centering\arraybackslash}m{0.6cm} >{\centering\arraybackslash}m{0.6cm} >{\centering\arraybackslash}m{0.6cm} >{\centering\arraybackslash}m{0.6cm} >{\centering\arraybackslash}m{0.6cm} >{\centering\arraybackslash}m{0.6cm}} 
			\toprule
			\multirow{2}{*}{$\alpha$} & \multicolumn{10}{c}{$n$}\\ [1ex]
			& $5$ & $6$ & $7$ & $8$ & $9$ & $10$ & $20$ & $30$ & $40$ & $50$\\ 
			\midrule
			\multicolumn{11}{c}{$D_n$: Monte Carlo vs. Algorithm \ref{alg:p-val}} \\ [1ex]
			$0.05$ & $14$ & $16$ & $19$ & $23$ & $16$ & $28$ & $69$ & $118$ & $182$ & $261$ \\ 
			\midrule
			\multicolumn{11}{c}{$W^2_n$: Monte Carlo vs. Algorithm \ref{alg:p-val}} \\ [1ex]
			$0.05$ & $10$ & $12$ & $14$ & $13$ & $17$ & $21$ & $51$ & $94$ & $146$ & $203$ \\ 
			\midrule
			\multicolumn{11}{c}{$A^2_n$: Monte Carlo vs. Algorithm \ref{alg:p-val}} \\ [1ex]
			$0.05$ & $15$ & $19$ & $22$ & $26$ & $32$ & $33$ & $80$ & $150$ & $227$ & $325$ \\ 
			\bottomrule
	\end{tabular}
	\fi
	\caption{\small Running time ratios, in scale $\times 10^4$, between a $p$-value Monte Carlo approximation based on $10^4$ trials and our $p$-value approximation method (Algorithm \ref{alg:p-val}). Ratios are computed for the median running times of $10^3$ evaluations, for each pair $(n, \alpha)$. The averages of the median running times for the Monte Carlo approximation are $2.34 s$, $2.35 s$, and $2.35 s$ for $D_n$, $W^2_n$, and $A^2_n$, respectively.}
	\label{table:exec-timeMC}
\end{table}

The empirical results show that our stabilized statistics give more accurate results than those by Stephens, while still retaining the simplicity of the latter. When it comes to the Monte Carlo approximation (with $10^4$ trials), relative errors on the empirical rejection rates are lowered by a factor that varies from $\times4$ to $\times10$, depending on the statistic. In addition, Table \ref{table:exec-timeMC} shows how our stabilization algorithm outperforms Monte Carlo execution times. Part of these improvements could be attributed to the R-C++ mix, as opposed to pure C++. Yet, given the massive difference in timing orders, we regard this effect as second-order. Arguably, for $D_n$, $W_n^2$, and $A_n^2$, the tailored approximation methods are to be preferred due to their better accuracy. Even in these highly-competitive settings, our stabilizations still offer comparative advantages, as Figure \ref{fig:rejection-proportion-MC-error} shows that their average relative error is $<0.7\%$, sufficing for most practical applications, while Table \ref{table:exec-time} shows an improvement of $\times5$ in running times with respect to specific methods.

\section{Stabilization of parameter-dependent statistics}
\label{section:parameter-modification}

This section gives an extension of the $(n, \alpha)$-transformations introduced in Section \ref{section:n-alpha-mod} that is designed to stabilize the exact distributions of statistics that depend on a (known) parameter. Instances of the transformation are given for testing uniformity on $\mathbb{S}^{p-1}$, $p\geq2$ being the statistic parameter.

\subsection{Projected-ecdf test statistics}
\label{section:proj-based-unif-tests}

\cite{Garcia-Portugues2020b} proposed a class of test statistics to evaluate the null hypothesis of uniformity of an iid sample $\mathbf{X}_1,\ldots,\mathbf{X}_n$ on $\mathbb{S}^{p-1}$. Projected-ecdf statistics compute the weighted quadratic discrepancy between $F_{n,\boldsymbol{\gamma}}$, the ecdf of $\boldsymbol{\gamma}'\mathbf{X}_1,\ldots,\boldsymbol{\gamma}'\mathbf{X}_n$ for $\boldsymbol{\gamma}\in\mathbb{S}^{p-1}$, and $F_p$, the cdf of the random variable $\boldsymbol{\gamma}'\mathbf{X}$ when $\mathbf{X}\sim\operatorname{Unif}(\mathbb{S}^{p-1})$. The weighted quadratic discrepancies are integrated over all possible directions $\boldsymbol{\gamma}\in\mathbb{S}^{p-1}$, a convenient specification of the projected-ecdf statistics being
\begin{align*}
	P_{n, p}^w : = n\int_{\mathbb{S}^{p-1}}\bigg[\int_{-1}^{1}\left(F_{n, \boldsymbol{\gamma}}(x) - F_{p}(x)\right)^2 w(F_p(x))\,\mathrm{ d}F_{p}(x)\bigg]\,\mathrm{d}\boldsymbol{\gamma},
\end{align*}
where $w:[0, 1]\rightarrow\mathbb{R}$ is a certain weight function and the cdf $F_p$ is that of the random variable $T$, with $T^2\sim\mathrm{Beta}(1/2,(p-1)/2)$.

The weights $w\equiv1$ and $w(u)=1/(u(1-u))$ result in the Projected Cramér--von Mises statistic, $P_{n, p}^{\CvM}$, and the Projected Anderson--Darling statistic, $P_{n, p}^{\AD}$, respectively. The test based on $P_{n, p}^{\CvM}$ happens to be an extension of the Watson test to $\mathbb{S}^{p-1}$, $p\geq2$, since $P_{n, 2}^{\CvM} = U^2_n/2$. Moreover, the test based on $P_{n, 3}^{\CvM}$ is equivalent to the chordal-based test on $\mathbb{S}^{2}$ by \cite{Bakshaev2010}, whose statistic for $p\geq 2$ is 
\begin{align*}
	N_{n,p}:=n\mathbb{E}_{\mathcal{H}_0}\left[\|\mathbf{X}_1-\mathbf{X}_2\|\right]-\frac{1}{n}\sum_{i,j=1}^n \|\mathbf{X}_i-\mathbf{X}_j\|.
\end{align*}
The statistic $P_{n, p}^{\AD}$ represents the first instance of the Anderson--Darling statistic in the context of directional data. Particularly, $P_{n, 2}^{\AD}$ can be regarded as the circular variant of $A_n^2$, just as $U_n^2$ is the circular variant of $W_n^2$. Asymptotic distributions and computational formulae for $P_{n, p}^{\CvM}$ and $P_{n, p}^{\AD}$ are provided in \cite{Garcia-Portugues2020b}, while the \texttt{sphunif} R package \citep{ Garcia-Portugues2020c} provides implementations for $P_{n, p}^{\CvM}$, $P_{n, p}^{\AD}$, and $N_{n,p}$, for all ${p\geq2}$.

\subsection{Stabilization of projected-ecdf statistics}
\label{section:modification-projected-ecdf}

Let $T_{n, p}$ be a statistic depending on $p\in\mathbb{N}$. From expression \eqref{eq:f_n_alpha}, the ratios $T_{\infty, p; \alpha} / T_{n, p; \alpha}$ can be modeled as a function $g:\mathbb{N}\times\mathbb{N}\times(0,1)\to\mathbb{R}$ of $(n,p,\alpha)$. Hence, the modified version of the statistic $T_{n, p}$ is defined as 
\begin{align*}
	T_{n,p}^{\ast}(\alpha) := T_{n, p} \cdot g(n, p,\alpha).
\end{align*}
As in expression \eqref{eq:reg}, the stabilization of $T_{n,p}$ can be approached as a regression problem, now with predictors $(n, p, \alpha)$, response $Y:=T_{\infty, p; \alpha} / T_{n, p; \alpha}$, and unknown regression function $g$.

The connection between $P_{n, 2}^{\CvM}$ and $U_{n}^2$ implies the stabilized form of $P_{n, 2}^{\CvM}$ to have the same set of predictors based on $(n, \alpha)$ as the Watson statistic already presented in Table \ref{table:modified-stats}: $\mathcal{R} := \{1/n, 1/(n\sqrt{\alpha}), 1/(n\alpha)\}$. An additional reflection suggests the adequacy of choosing $\mathcal{R}$ for stabilizing $P_{n, p}^{\CvM}$, also when $p\geq 3$, due to its appearance in all the transformations for quadratic-ecdf statistics in Table \ref{table:modified-stats} and the quadratic nature of $P_{n, p}^{\CvM}$. For different particular values of $p\geq 2$, it was noted that, if regression models were fitted to the ratios $P_{\infty, p; \alpha}^{\CvM}/P_{n, p; \alpha}^{\CvM}$, the coefficients fitted for each predictor $r \in \mathcal{R}$ could be modeled as a smooth function of $p$ denoted as $q_r:\mathbb{N}\to\mathbb{R}$. Unsurprisingly, given its similarity to $P_{n, p}^{\CvM}$, the same considerations also hold for $P_{n, p}^{\AD}$. Moreover, the statistic $N_{n,p}$ can also be stabilized through $\mathcal{R}$ and $q_r$, a fact explained by the closeness between $P_{n, p}^{\CvM}$ and $N_{n,p}$ when $p\neq 3$ and its equivalence when $p=3$. Empirical investigations suggested the following saturated model for $q_r$, for each $r \in \mathcal{R}$:
\begin{align*}
	q_r(p) = \frac{\beta_{r,1}}{\sqrt{p}} + \frac{\beta_{r,2}}{p}.
\end{align*}
Thus, the resulting saturated model for $g$ is set as
\begin{align}
	g(n, p, \alpha) = 1 + q_{1/n}(p)\cdot\frac{1}{n} + q_{1/(n\alpha)}(p)\cdot\frac{1}{n\alpha} 
	+ q_{1/(n\sqrt{\alpha})}(p)\cdot\frac{1}{n\sqrt{\alpha}}. \label{eq:saturated_g_n_p_alpha}
\end{align}

Once training samples of the form $\{(n_j,\alpha_j,p_j,Y_j)\}_{j=1}^J$, $Y_j:=T_{\infty, p_j; \alpha_j} /\allowbreak T_{n_j, p_j; \alpha_j}$, are available, model \eqref{eq:saturated_g_n_p_alpha} is estimated following the same methodology described in Section \ref{section:n-alpha-mod}. For each of the three test statistics $P_{n,p}^{\CvM}$, $P_{n,p}^{\AD}$, and $N_{n,p}$, we obtained $M = 10^7$ Monte Carlo random samples for each sample size $n$ in $\mathcal{N}:=\{5,\ldots,100, 102,\ldots, 200, 204, \ldots,\allowbreak300,308,\ldots,\allowbreak404,420,\allowbreak\ldots,500\}$ and for each dimension $p$ in $\mathcal{P}:=\{2,\ldots,\allowbreak11,21,31,41,\allowbreak51,61,71,\allowbreak81,91,\allowbreak101,151,201,251,\allowbreak301\}$. We then summarized these statistics as the quantiles $\{T_{n_j,p_j;\alpha_j}: n_j\in\mathcal{N},\,p_j\in\mathcal{P},\,\alpha_j\in\mathcal{A}\}$ for $\mathcal{A}:=\{a/A:a=1,\ldots,A\}$, $A=10^3$. The asymptotic $\alpha$-quantiles $T_{\infty, p; \alpha}$ were approximated through $T_{500,p;\alpha}$ due to the accuracy limitations on inverting the asymptotic cdfs of the three statistics for large dimensions. Table \ref{table:modified-sph-stats} lists the approximated $T_{\infty,p;\alpha}$ for the first ten dimensions. 

\begin{table}[ht!]
	\iffigstabs
	\centering
	\small
	\begin{tabular}{>{\centering\arraybackslash}m{1.5cm} >{\centering\arraybackslash}m{1cm} >{\centering\arraybackslash}m{0.8cm} >{\centering\arraybackslash}m{0.8cm} >{\centering\arraybackslash}m{0.8cm} >{\centering\arraybackslash}m{0.8cm} >{\centering\arraybackslash}m{0.8cm} >{\centering\arraybackslash}m{0.8cm} >{\centering\arraybackslash}m{0.8cm} >{\centering\arraybackslash}m{0.8cm} >{\centering\arraybackslash}m{0.8cm} >{\centering\arraybackslash}m{0.8cm}} 
			\toprule
			& & \multicolumn{10}{c}{$p$}\\ [1ex]
			Critical value & $\alpha$ & $2$ & $3$ & $4$ & $5$ & $6$ & $7$ & $8$ & $9$ & $10$ & $11$ \\ 
			\midrule
			\multirow{5}{*}{$P_{\infty,p;\alpha}^{\CvM}$}
			& $0.10$ & $0.3035$ & $0.2768$ & $0.2606$ & $0.2500$ & $0.2421$ & $0.2361$ & $0.2312$ & $0.2272$ & $0.2239$ & $0.2210$ \\ [1ex]
			& $0.05$ & $0.3735$ & $0.3288$ & $0.3027$ & $0.2858$ & $0.2735$ & $0.2641$ & $0.2568$ & $0.2508$ & $0.2458$ & $0.2416$ \\ [1ex]
			& $0.01$ & $0.5358$ & $0.4461$ & $0.3960$ & $0.3638$ & $0.3413$ & $0.3244$ & $0.3115$ & $0.3008$ & $0.2922$ & $0.2849$ \\ [1ex]
			\midrule
			\multirow{5}{*}{$P_{\infty,p;\alpha}^{\AD}$}
			& $0.10$ & $1.6871$ & $1.5604$ & $1.4816$ & $1.4279$ & $1.3883$ & $1.3576$ & $1.3327$ & $1.3124$ & $1.2957$ & $1.2809$ \\ [1ex]
			& $0.05$ & $2.0293$ & $1.8214$ & $1.6951$ & $1.6106$ & $1.5494$ & $1.5023$ & $1.4651$ & $1.4347$ & $1.4092$ & $1.3875$ \\ [1ex]
			& $0.01$ & $2.8197$ & $2.4096$ & $2.1679$ & $2.0090$ & $1.8969$ & $1.8126$ & $1.7471$ & $1.6931$ & $1.6493$ & $1.6121$ \\ [1ex]
			\midrule
			\multirow{5}{*}{$N_{\infty,p;\alpha}$}
			& $0.10$ & $2.4034$ & $2.2141$ & $2.1003$ & $2.0231$ & $1.9673$ & $1.9238$ & $1.8887$ & $1.8601$ & $1.8367$ & $1.8158$ \\ [1ex]
			& $0.05$ & $2.9906$ & $2.6305$ & $2.4320$ & $2.3034$ & $2.2119$ & $2.1423$ & $2.0879$ & $2.0437$ & $2.0067$ & $1.9752$ \\ [1ex]
			& $0.01$ & $4.3495$ & $3.5687$ & $3.1669$ & $2.9136$ & $2.7402$ & $2.6112$ & $2.5124$ & $2.4314$ & $2.3661$ & $2.3108$ \\ [1ex]
			\bottomrule
	\end{tabular}
	\fi
	\caption{\small Asymptotic critical values for modified uniformity statistics with dimension $p$, sample size $n$, and significance level $\alpha$.}
	\label{table:modified-sph-stats}
\end{table}

\begin{table}[ht!]
	\iffigstabs
	\centering
	\small
	\begin{tabular}{>{\centering\arraybackslash}m{1.5cm} >{\centering\arraybackslash}m{2.3cm} >{\centering\arraybackslash}m{2.3cm} >{\centering\arraybackslash}m{2.3cm}} 
		\toprule
		\multirow{2}{*}{$T_{n,p}$} & \multicolumn{3}{c}{$T_{n, p} \left(1 + q_{1/n}\cdot\frac{1}{n} + q_{1/(n\alpha)}\cdot\frac{1}{n\alpha} + q_{1/(n\sqrt{\alpha})}\cdot\frac{1}{n\sqrt{\alpha}}\right)$}\\ [1ex]
		& $q_{1/n}$ & $q_{1/(n\alpha)}$ & $q_{1/(n\sqrt{\alpha})}$\\
		\midrule
		$P_{n,p}^{\CvM}$ & $\frac{0.1130}{\sqrt{p}} - \frac{0.5415}{p}$ & $- \frac{0.0031}{\sqrt{p}}$ & $\frac{0.1438}{\sqrt{p}}$ \\ [3ex]
		$P_{n,p}^{\AD}$ & $\frac{0.0978}{\sqrt{p}} -\frac{0.3596}{p}$ & $- \frac{0.0025}{\sqrt{p}}$ & $\frac{0.1126}{\sqrt{p}}$ \\ [3ex]
		$N_{n,p}$ & $\frac{0.1189}{\sqrt{p}} - \frac{0.5838}{p}$ & $- \frac{0.0030}{\sqrt{p}}$ & $\frac{0.1210}{\sqrt{p}} + \frac{0.0385}{p}$ \\ [1.5ex]
		\bottomrule
	\end{tabular}
	\fi
	\caption{\small Modified uniformity statistics for dimension $p$, sample size $n$, and significance level $\alpha$. Modified forms are valid for $2\leq p \leq 300$, $n \ge 5$, and $\alpha \leq 0.25$. $\mathcal{H}_0$ is rejected at significance level $\alpha$ if $T_{n,p}^{\ast}(\alpha)>T_{\infty,p;\alpha}$, where $T_{\infty,p;\alpha}$ is given in Table \ref{table:modified-sph-stats} for $p=2,\ldots,11$.}
	\label{table:modified-forms-sph-stats}
\end{table}

The resulting modified forms for $P_{n,p}^{\CvM}$, $P_{n,p}^{\AD}$, and $N_{n,p}$ are presented in Table \ref{table:modified-forms-sph-stats}, where each fitted $q_r$ is shown for each predictor $r\in \mathcal{R}$. An algorithm similar to Algorithm \ref{alg:p-val} for computing an approximated $p$-value has been implemented for these statistics, with the only difference being that the modified statistic function in lines $3$ and $11$ is the corresponding dimension-dependent version which also includes the parameter $p$ as an input.

\subsection{Simulation study}
\label{section:simulations2}

In the same manner as in Section \ref{section:simulations}, the empirical stabilization of the modified forms of the projected-ecdf statistics is investigated (Figure \ref{fig:error-projbased-statistics}) in terms of the relative error between the significance level and the empirical rejection rate of the $T_{n,p}^\ast(\alpha)$-test for sample sizes $n \in \mathcal{N}_{\text{test}}$ and dimensions $p \in \mathcal{P}_{\text{test}}$, where $\mathcal{N}_{\text{test}}$ was defined in Section \ref{section:simulations} and $\mathcal{P}_{\text{test}} := \{2, \ldots, 11, 21, 51, 101, 151,\allowbreak 201,\allowbreak 301\}$. As for most non-heavily studied test statistics, Monte Carlo is the only method readily available to approximate the exact-$n$ $p$-values of $P_{n,p}^{\CvM}$, $P_{n,p}^{\AD}$, and $N_{n,p}$. Figure \ref{fig:error-projbased-statistics} shows an average improvement of our stabilizations' accuracy over Monte Carlo approximations (using $10^4$ trials) of $\times3$, $\times4$, and $\times4$, for each of the three statistics, respectively. We point out the steadiness of our relative errors regardless of the significance level and the dimension $p$ (except for $\alpha = 0.01$, which increases for large $p$'s), which on average are $1.3\%$, $0.9\%$, and $1\%$ respectively. In almost all circumstances, our relative errors are largely below those obtained by Monte Carlo (except for $\alpha=0.25$ when $p>10$ in $P_{n,p}^{\CvM}$ and $N_{n,p}$).

\begin{figure}[ht!]
	\iffigstabs
	\centering
	\begin{subfigure}[t]{0.5\textwidth}
		\includegraphics[width=\linewidth]{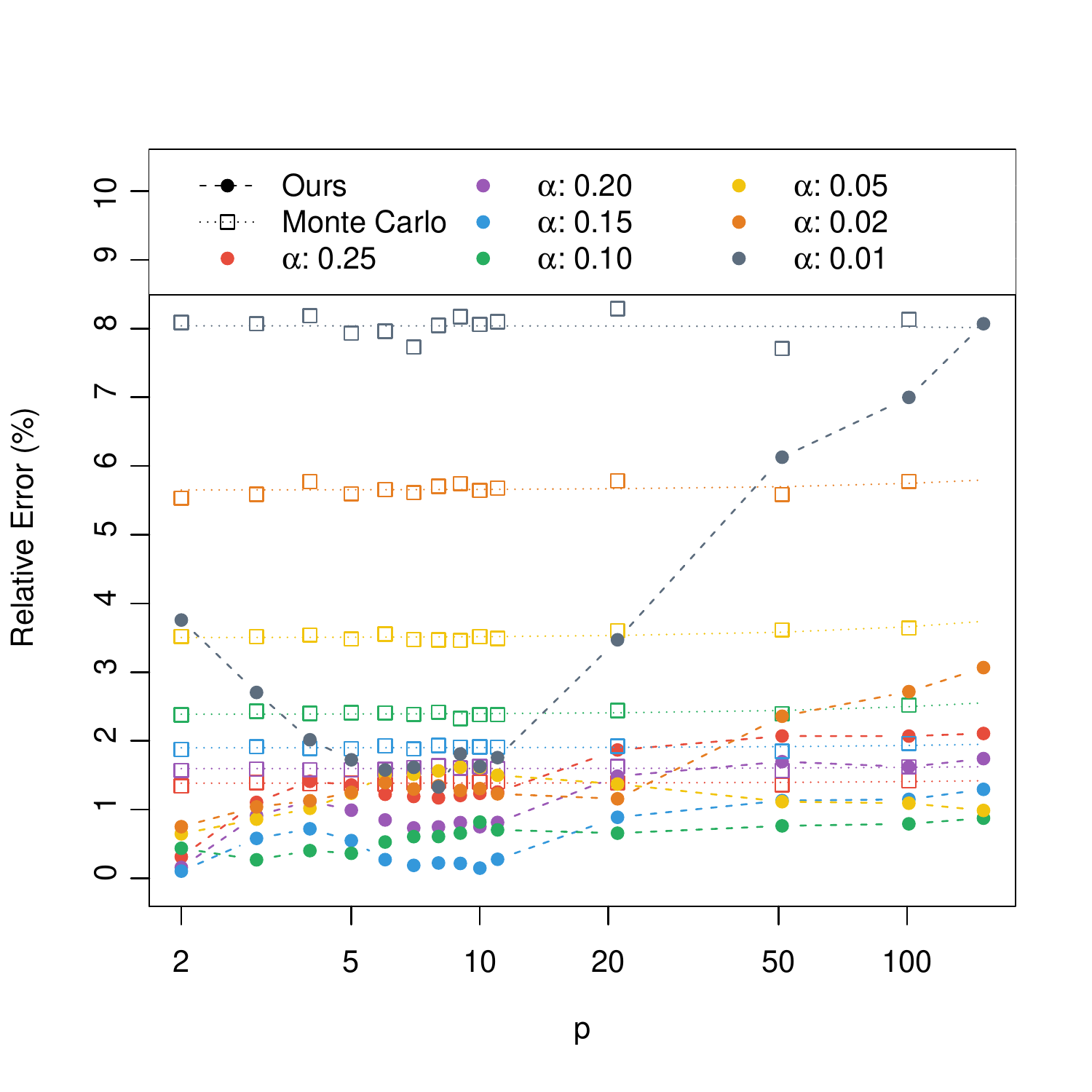}
		\caption{\small $P_{n,p}^{\CvM}$\label{fig:CvM}}
	\end{subfigure}%
	\begin{subfigure}[t]{0.5\textwidth}
		\includegraphics[width=\linewidth]{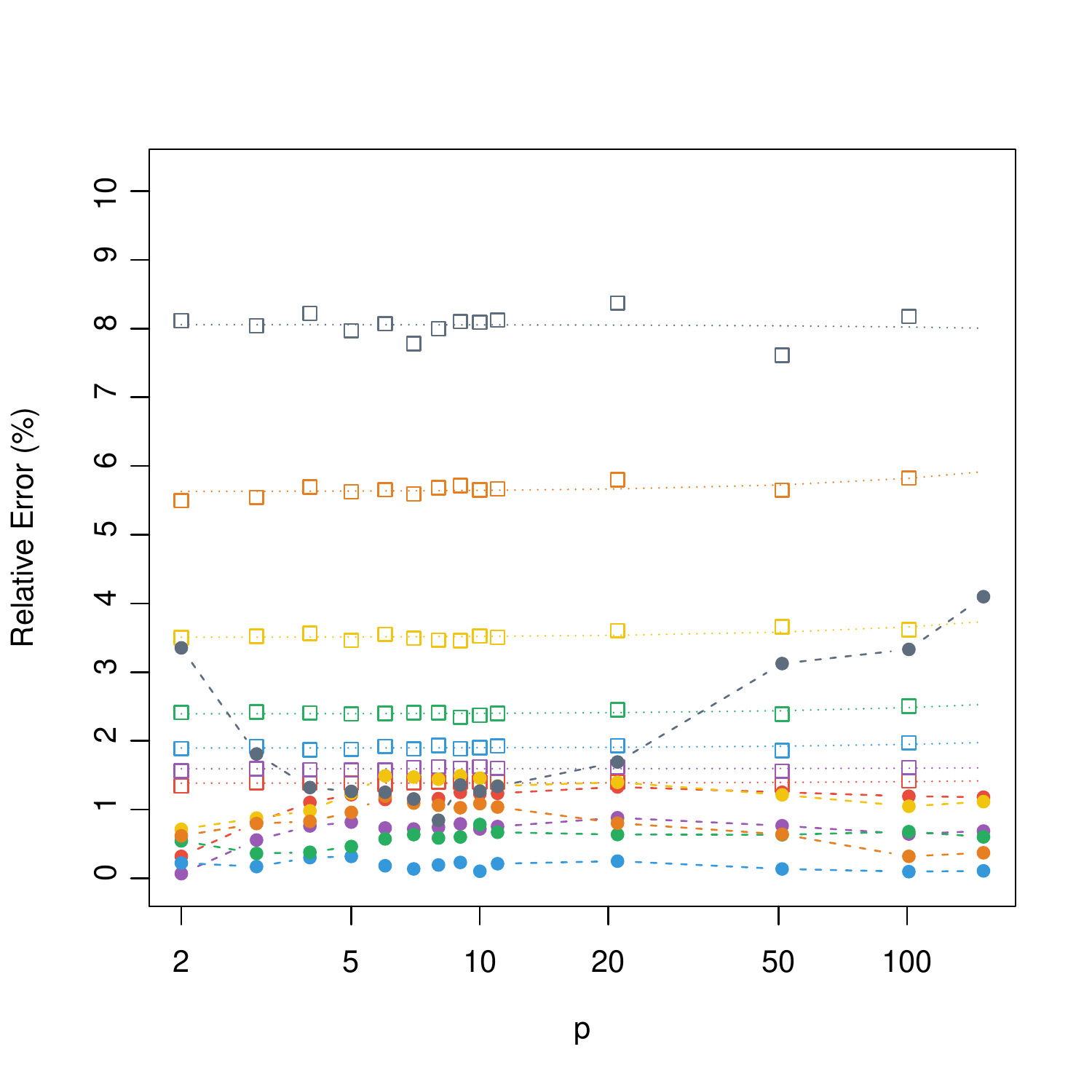}
		\caption{\small $P_{n,p}^{\AD}$}
	\end{subfigure}
	\begin{subfigure}[t]{0.5\textwidth}
		\includegraphics[width=\linewidth]{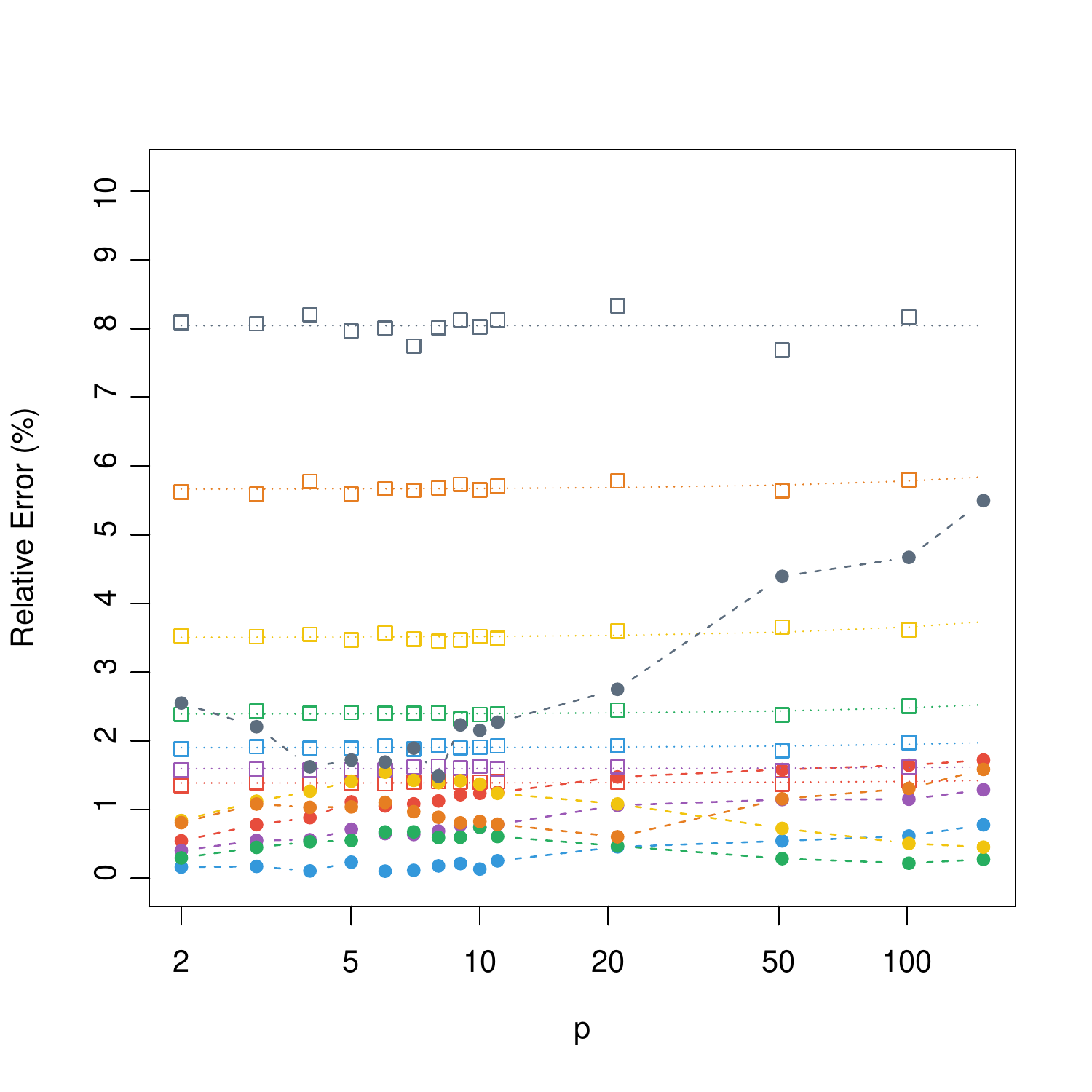}
		\caption{\small $N_{n,p}$}
	\end{subfigure}
	\fi
	\caption{\small Relative error (in \%) $\lvert\alpha-\tilde{\alpha}\rvert/\alpha$ between the significance level $\alpha$ and $\tilde{\alpha}$, the empirical rejection rate using an approximated exact-$n$ critical value, averaged over $5\leq n \leq 300$. For the Monte Carlo approximation method, a regression fit is shown for each significance level $\alpha$ to show no trend on the error with respect to $p$. The legend in Figure \ref{fig:CvM} details the approximation methods considered and significance levels, and applies to the rest of the panels.}
	\label{fig:error-projbased-statistics}
\end{figure}

We conclude this section by summarizing in Table \ref{table:summary-table} a comparison of the modified forms found by \cite{Stephens1970} and our results, for each of the classical ecdf-based statistics, and their corresponding versions for circular data, along with the circular particularizations of the projected-ecdf statistics. We emphasize the simplicity of the formulae in both approaches, with the Mean Relative Error (MRE) being reduced for the second by $\times2$ for $D_n$ and $U_n^2$, by $\times9$ for $W_n^2$, and by $\times4$ for $A_n^2$ and $V_n$. The stabilizations for the projected-ecdf statistics are such $\text{MRE}<0.9\%$ for the circular case, which aligns with the results specifically attained for $U_n^2$ and $P_{n,2}^{\AD}$, and supports the convenience of the extension proposed in this section for the $(n, \alpha)$-stabilization.

\begin{table}[ht!]
	\iffigstabs
	\centering
	\small
		\begin{tabular}{>{\centering\arraybackslash}p{1.8cm} >{\centering\arraybackslash}p{4cm} >{\centering\arraybackslash}p{1cm} >{\centering\arraybackslash}p{5.2cm} >{\centering\arraybackslash}p{1cm}}
			\toprule
			$T_n$ & Stephens' $T_n^{\ast}$ & MRE & $T_n^{\ast}(\alpha)$ & MRE \\
			\midrule
			$D_n$ & $D_n\left(1 + \frac{0.12}{\sqrt{n}} + \frac{0.11}{n}\right)$ & $1.44\%$ & $D_n\left(1 + \frac{0.1575}{\sqrt{n}} + \frac{0.0192}{n\sqrt{\alpha}} - \frac{0.0051}{\sqrt{n\alpha}}\right)$ & $0.63\%$\\ [3ex]
			$W_n^2$ & $\left(W_n^2-\frac{0.4}{n}+\frac{0.6}{n^2}\right)\left(1 + \frac{1}{n}\right)$ & $3.28\%$ & $W_n^2\left(1 - \frac{0.1651}{n} + \frac{0.0749}{n\sqrt{\alpha}} - \frac{0.0014}{n \alpha}\right)$ & $0.36\%$ \\ [3ex]
			$A_n^2$ & $A_n^2$ $(\ssymbol{1})$ & $1.42\%$ & $A_n^2\left(1 + \frac{0.0360}{n} - \frac{0.0234}{n \sqrt{\alpha}} + \frac{0.0006}{n \alpha}\right)$ & 0.38\%\\ [3ex]
			$V_n$ & $V_n\left(1 + \frac{0.155}{\sqrt{n}} + \frac{0.24}{n}\right)$ & $3.40\%$ & $V_n\left(1 + \frac{0.2330}{\sqrt{n}} + \frac{0.0276}{n\sqrt{\alpha}} - \frac{0.0068}{\sqrt{n\alpha}}\right)$ & $0.85\%$ \\ [3ex]
			$U_n^2 \equiv P^{\CvM}_{n, 2}$ & $\left(U_n^2-\frac{0.1}{n}+\frac{0.1}{n^2}\right)\left(1 + \frac{0.8}{n}\right)$ & $1.62\%$ & $U_n^2\left(1 - \frac{0.1505}{n} + \frac{0.0917}{n\sqrt{\alpha}} - \frac{0.0018}{n \alpha}\right)$ & $0.63\%$ \\ [3ex]
			& $-$ & $-$ & $P^{\CvM}_{n, 2}\left(1 - \frac{0.1908}{n} + \frac{0.1017}{n\sqrt{\alpha}} - \frac{0.0022}{n\alpha}\right)$ & $0.88\%$ \\ [3ex]
			$P^{\AD}_{n, 2}$ $(\ssymbol{2})$ & $-$ & $-$ & $P^{\AD}_{n, 2} \left(1 - \frac{0.0751}{n} + \frac{0.0692}{n \sqrt{\alpha}} - \frac{0.0014}{n \alpha}\right)$ & $0.74\%$\\ [3ex]
			& $-$ & $-$ & $P^{\AD}_{n, 2} \left(1 - \frac{0.1106}{n} + \frac{0.0796}{n \sqrt{\alpha}} - \frac{0.0018}{n \alpha}\right)$ & $0.83\%$\\ [3ex]
			\bottomrule
	\end{tabular}
	\fi
	\caption{\small Modified forms of ecdf-based statistics for sample size $n$ and significance level $\alpha$. MRE refers to the Mean Relative Error between the expected rejection proportion and the approximated proportion for each pair of $(n, \alpha)$ with $n \in \mathcal{N}_{\text{test}}$ and $\alpha \in \{0.25, 0.2, 0.15, 0.1, 0.05, 0.02, 0.01\}$. The $T_n^\ast(\alpha)$ forms are valid for $n \ge 5$ and $\alpha \leq 0.25$. $(\ssymbol{1})$ \cite{Stephens1974} states the best modification for Anderson--Darling statistic for $n \geq 5$ is its asymptotic distribution. $(\ssymbol{2})$ Both the modified form estimated for $p = 2$ (top row) and the $(n, p, \alpha)$-modification particularized for $p = 2$ (bottom row) are given for $P^{\CvM}_{n, 2}$ and $P^{\AD}_{n, 2}$.}
	\label{table:summary-table}
\end{table}

\section{Detecting temporal longitudinal non-uniformity in sunspots}
\label{section:applications}

The Sun's magnetic field presents periodic behavioral patterns of about $11$ years. During this period, the magnetic field is pulled around the Sun's surface as the solar plasma rotates. Near the equator this pull is stronger due to the Sun's differential rotation (equatorial latitudes rotate faster than poles), causing the field to wrap in a spiral-like shape until its polarity is eventually reversed and the wrapping restarts, indicating the beginning of a new solar cycle \citep[see, e.g.,][]{Babcock1961}. Sunspots are created by high-intensity magnetic loops emerging from the Sun's interior convection zone to the surface, producing darker, cooler regions on the Sun's photosphere. Through their lifespans, which can last from hours to days, they experience continuous changes in shape, area, and location. The total number of active sunspots varies throughout the cycle, showing the maximum activity during the middle years (see Figure \ref{fig:sunspots}). Sunspots appear in a marked rotationally symmetric fashion: they are mainly distributed in latitudinal belts that are initially situated at $\pm30^{\circ}$ and that decay to the equator as the solar cycle advances (a phenomenon known as the \textit{Sp\"orer's law}).

Sunspots also appear to cluster in \textit{active longitudes}. Non-uniform patterns may appear by \textit{preferred zones of occurrence} where sunspots had originated previously, as early described by \citet[pages 574 and 581]{Babcock1961}. The existence of active longitudes was also suggested in \cite{Bogart1982} upon inspection of the significant autocorrelation of daily sunspot numbers. Since daily sunspot numbers have no positional information, such analysis shows either there is one active longitude band or there are two active longitude bands separated by $180^\circ$, as \cite{Berdyugina2003} concluded in both hemispheres, observing the alternation of major solar activity between both longitudes in about $1.5$ to $3$ years. This is known as the \textit{flip-flop} phenomenon \citep{Elstner2005}.

Analyzing the presence of solar active longitudes requires knowledge from the Carrington period (or solar rotation period). It corresponds to the mean synodic rotation period of sunspots, which is about $27.275$ days. Differential rotation causes the migration of active longitudes in the Carrington reference frame, causing a lag of 2.5 Carrington rotations per solar cycle that blurs the longitudinal pattern if more than one solar cycle is analyzed at once \citep{Berdyugina2003}. In order to ensure the adequate detection of active longitudes, a sequential analysis of data limited to a certain number of Carrington rotations, from 3--7 \citep{Bogart1982, deToma2000} to 10--15 \citep{Pelt2010}, is preferable.

The data we analyze is based on the Debrecen Photoheliographic Data (DPD) sunspot catalog \citep{Baranyi2016, Gyoeri2016}. It contains observations of sunspots locations since 1974 and is a continuation of the Greenwich Photoheliographic Results (GPR) catalog, which spanned 1872--1976. The dataset $\texttt{sunspots\_births}$, available in the R package \texttt{rotasym} \citep{Garcia-Portugues2020e}, accounts just for the first-ever observation (referred to as ``birth'' henceforth) of a group of sunspots.

In our analysis, summarized in Figure \ref{fig:sunspots}, we first applied the $P_{n, 2}^{\AD}$-based uniformity test sequentially to the longitudes of sunspot births ---which include a total of $6195$, $4551$, and $5373$ observations for the 21st, 22nd, and 23rd cycle, respectively--- within a rolling window whose size is $10$-Carrington rotations (approximately, nine months). The corresponding $p$-values were computed using Algorithm \ref{alg:p-val} for northern (blue), southern (red), and both (black) hemispheres. In addition, the $p$-value was also computed by Monte Carlo with $5\times10^3$ samples, in order to compare the running times between the two methods. Our method runs in an average of $1.6\text{ s}$ per solar cycle, while Monte Carlo completes it in $1600\text{ s}$ per solar cycle. In order to account for dependency between sequential tests, \cite{Benjamini2001}'s FDR correction was applied to the $p$-values obtained with the test based on $P_{n, 2}^{\AD}$. These corrected $p$-values are shown in the top row of Figure \ref{fig:sunspots}. Second, circular-linear kernel density estimation \citep{Garcia-Portugues2013b} of sunspot births for northern (middle-top figure) and southern (middle-bottom) hemispheres allowed us to compute several level sets, represented as contour lines labeled as ``$100p \%$''. Each of these sets is the smallest set containing $1-p$ of the probability of the estimated density function. Hence, darker sets represent higher-density zones of sunspot births, both through time and longitude. Third, a scatter plot of sunspot births is shown in the bottom figures, along with the circular Nadaraya--Watson \citep{DiMarzio2012} regression for northern (blue), southern (red), and both (black) hemispheres. The Nadaraya--Watson regression gives a moving circular mean of the longitudes of sunspot births through time. Both density and regression kernel estimates use ``rule-of-thumb'' bandwidths for normal \citep{Silverman1986} and von Mises--Fisher \citep{Garcia-Portugues2013a} distributions, given the similarity of marginal distributions with these respective distributions and the marked undersmoothing that resulted from cross-validation bandwidths.

\begin{sidewaysfigure}
	\iffigstabs
	\centering
	\includegraphics[width=1\linewidth]{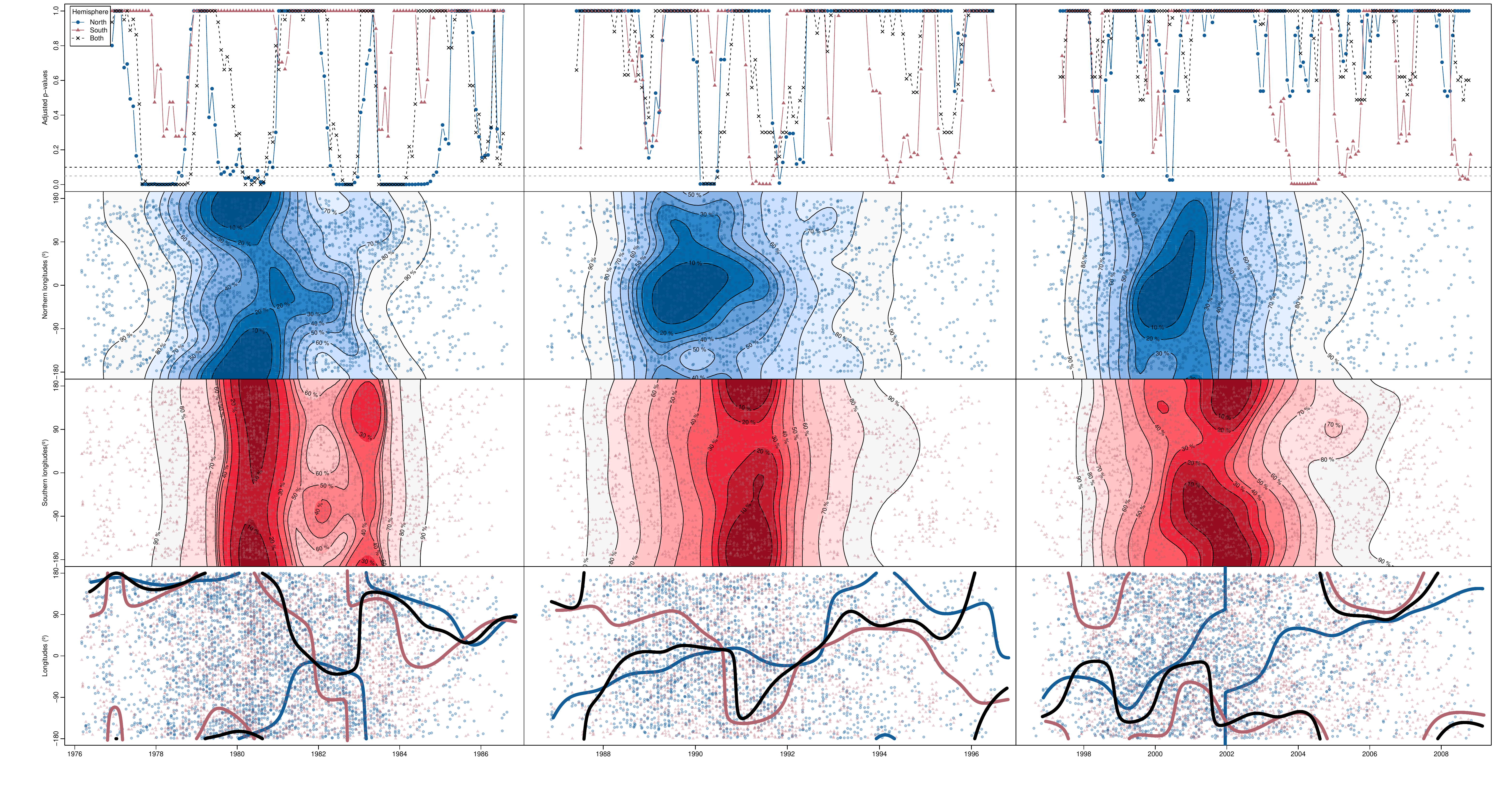}
	\fi
	\caption{\small Longitudinal non-uniformity patterns of sunspot births. Each column represents the analysis for each of the 21st, 22nd, and 23rd solar cycles. Northern (blue), southern (red), and both (black) hemispheres were separately analyzed. Top figures: $P_{n, 2}^{\AD}$-based uniformity test of sunspot births longitudes. The $p$-values shown are corrected by \cite{Benjamini2001}'s FDR. Middle figures: Circular-linear kernel density level sets of sunspot births through time and longitude. Bottom figures: Sunspot births (points) along with the corresponding Nadaraya--Watson regression (lines).}
	\label{fig:sunspots}
\end{sidewaysfigure}

We draw the following conclusions from the analysis:
\begin{enumerate}[label=(\textit{\roman*}),ref=(\textit{\roman*})]
	\item In general, the two hemispheres seem to have different behavioral patterns, both in terms of longitudinal non-uniformities and sunspots activity level, along solar cycles. During the 21st cycle, the northern hemisphere presents $33\%$ of the tests rejected at significance level $\alpha=0.05$. In cycles 22 and 23, the southern hemisphere presents more non-uniform periods ($9\%$ and $10\%$ of the tests are rejected for $\alpha=0.05$, respectively) than the northern hemisphere ($5\%$ and $3\%$ are rejected, respectively).
	
	\item Non-uniformity periods are intermittent during the lifetime of the solar cycle, without a clear association with the intensity of the sunspots appearance. The length and quantity of non-uniformity periods differ between solar cycles.
	
	\item Sunspots seem to appear in preferred zones of occurrence. Highest density sets, together with Nadaraya--Watson regressions, show consistent patterns of activity within certain longitudinal zones. In particular, periods in which uniformity is rejected at significance level $\alpha=0.05$, the northern sunspot births seem to cluster around $0^\circ$ (1982, 1990, 2000), $135^\circ$ (1983--1984), and $180^\circ$ (1977--1978, 1979--1980), while the southern hemisphere sunspot births cluster around $-135^\circ$ (1991, 2004, 2008). However, non-uniformity periods are too few to claim the existence of active longitudes.
	
	\item The flip-flop phenomenon between $180^\circ$ active longitudes is not obvious throughout all the cycles. Although longitudes $0$ and $180^\circ$ seem to accumulate more sunspots in the northern hemisphere, the alternation between supplementary longitudes is not a clear, fixed-duration pattern.
\end{enumerate}

\section{Discussion}
\label{section:discussion}

We have presented a general, automated approach to construct simple yet effective approximations for the upper tail of the exact-$n$ null distribution of numerous goodness-of-fit statistics. The simulation results demonstrate that these approximations are accurate enough for practical applications of several upper-tail tests, even when these depend on a varying (yet known) parameter.

Although state-of-the-art statistic-specific algorithms like those of \cite{Marsaglia2003}, \cite{Csorgo1996}, and \cite{Marsaglia2004} provide arbitrarily accurate upper-tail $p$-values for the $D_n$, $W_n^2$, and $A_n^2$ statistics, respectively, our $p$-value approximation method offers significant computational improvements, has a reasonable precision (mean relative errors below $1\%$), and, most importantly, can be applied to a wide range of statistics. Compared to the general and omnipresent $p$-value approximation by $M$ Monte Carlo trials, our method presents two key advantages: (\textit{i}) more accurate results (at least, when $M=10^4$); and (\textit{ii}) faster running times by several orders of magnitude. This computational expediency makes the stabilized statistic especially convenient for sequential tests, as illustrated in the data application.

The $(n, \alpha)$-stabilization significantly extends the scope of applicability of stabilizations like those of \cite{Stephens1970}. The stabilization focuses only on the upper tail of $T_n$, as this is usually the most useful in practice. However, stabilizations for the lower tail can be analogously derived. Obtaining modifications that stabilize the whole distribution, while still retaining simplicity, would offer the advantage of having approximated $p$-values that are roughly uniformly distributed under the null hypothesis. This task is left for future research.

\section*{Acknowledgments}

The second author acknowledges support by grants PGC2018-097284-B-100 and IJCI-2017-32005 by Spain's Ministry of Science, Innovation and Universities. The two grants were co-funded with ERDF funds. The computational resources of the Supercomputing Center of Galicia (CESGA) are greatly appreciated. The authors greatly acknowledge the comments of three referees.

\fi

\ifsupplement
\newpage

\title{Supplementary material for ``Data-driven stabilizations of goodness-of-fit tests''}
\setlength{\droptitle}{-1cm}
\predate{}
\postdate{}
\date{}

\author{Alberto Fern\'andez-de-Marcos$^{1,2}$ and Eduardo Garc\'ia-Portugu\'es$^{1}$}
\footnotetext[1]{Department of Statistics, Carlos III University of Madrid (Spain).}
\footnotetext[2]{Corresponding author. e-mail: \href{mailto:albertfe@est-econ.uc3m.es}{albertfe@est-econ.uc3m.es}.}
\maketitle

\begin{abstract}
    This supplementary material is divided into two sections. Appendix \ref{supplementary:powers} contains an analysis of regression metrics for different polynomial forms of \myeqref{eq:saturated_g_n_alpha}{6} in Section \myref{section:n-alpha-mod}{2.2}. Appendix \ref{supplementary:weights} presents an analysis of different weight functions for the regression in Section \myref{section:n-alpha-mod}{2.2}.
\end{abstract}

\appendix

\section{\texorpdfstring{Selection of $(n, \alpha)$-model form}{Selection of (n, alpha)-model form}}
\label{supplementary:powers}

In order to choose and justify the final form of model \myeqref{eq:saturated_g_n_alpha}{6}, an analysis of different models is presented in Tables \ref{table:powers-n} and \ref{table:powers-alpha}. The explored model specifications are
\begin{align}
    g_{\lambda, \mu}(n, \alpha) := 1 + \{\beta_{l, m}n^{-l/2}\alpha^{-m/2}\}_{l=1,m=0}^{\lambda,\mu}, \label{eq:glambdamu}
\end{align}
where $\lambda, \mu \in \mathbb{N}$ are to be tuned. With this notation, the model considered in \myeqref{eq:saturated_g_n_alpha}{6} in Section \myref{section:n-alpha-mod}{2.2} is $g_{2,2}$.

Table \ref{table:powers-n} shows a comparative study of the performance of $g_{\lambda, 2}(n, \alpha)$ for $\lambda=2, \ldots, 5$, in order to determine the optimal power of $n$ predictors, while keeping the $\alpha$-predictors unchanged. Conversely, Table \ref{table:powers-alpha} shows the performance of models $g_{2, \mu}(n, \alpha)$ where $\mu=2, \ldots, 5$ to analyze the effect of $\alpha$, while not varying the $n$-predictors. In both tables the same main model-selection procedure applied in Section \myref{section:n-alpha-mod}{2.2} is carried out: forward-backward model selection is run from Stephens' set of predictors, using weighted least squares with weights $\smash{w_j=n_j^{-1/2}1_{\{0<\alpha_j\leq 0.25\}}}$, and with extended scope given by \eqref{eq:glambdamu} (interactions are included). The last dropping step is excluded from the analysis.

\begin{table}[ht!]
	\iffigstabs
	\centering
	\small
	\scalebox{1}{
        \begin{tabular}{ m{0.4cm} m{0.2cm} wr{2cm} wr{1.3cm} wr{1cm} wr{1cm} wl{2cm}} 
        \toprule
        $T_n$ & $\lambda$ & $\hat{\sigma}$ & $\operatorname{BIC}$ & $R^2_{\mathrm{adj}}$ & $\operatorname{MVIF}$ & $\mathrm{pred}_{\operatorname{MVIF}}$\\ 
        \midrule
        \multirow{4}{0.6cm}{$D_n$} & $2$ & $7.57\cdot10^{-4}$ & $-397236$ & $0.9993$ & $1\cdot10^2$ & $n^{-1/2}\alpha^{-1/2}$ \\ 
        & $3$ & $7.50\cdot10^{-4}$ & $-398020$ & $0.9993$ & $4\cdot10^3$ & $n^{-1}\alpha^{-1/2}$ \\ 
        & $4$ & $7.48\cdot10^{-4}$ & $-398184$ & $0.9993$ & $1\cdot10^5$ & $n^{-3/2}\alpha^{-1/2}$ \\  
        & $5$ & $7.48\cdot10^{-4}$ & $-398212$ & $0.9993$ & $3\cdot10^5$ & $n^{-2}$ \\  
        \hline
        \multirow{4}{0.6cm}{$W^2_n$} & $2$ & $9.06\cdot10^{-4}$ & $-369813$ & $0.9835$ & $1\cdot10^2$ & $n^{-1/2}\alpha^{-1/2}$ \\ 
        & $3$ & $8.84\cdot10^{-4}$ & $-371092$ & $0.9841$ & $8\cdot10^2$ & $n^{-1}\alpha^{-1/2}$ \\ 
        & $4$ & $8.83\cdot10^{-4}$ & $-371112$ & $0.9841$ & $7\cdot10^3$ & $n^{-3/2}$ \\  
        & $5$ & $8.83\cdot10^{-4}$ & $-371119$ & $0.9841$ & $2\cdot10^5$ & $n^{-2}$ \\  
        \hline
        \multirow{4}{0.6cm}{$A^2_n$} & $2$ & $8.11\cdot10^{-4}$ & $-376072$ & $0.8722$ & $2\cdot10^1$ & $n^{-1}\alpha^{-1/2}$ \\ 
        & $3$ & $8.11\cdot10^{-4}$ & $-376072$ & $0.8722$ & $2\cdot10^1$ & $n^{-1}\alpha^{-1/2}$ \\ 
        & $4$ & $8.11\cdot10^{-4}$ & $-376072$ & $0.8722$ & $2\cdot10^1$ & $n^{-1}\alpha^{-1/2}$ \\ 
        & $5$ & $8.11\cdot10^{-4}$ & $-376072$ & $0.8722$ & $2\cdot10^1$ & $n^{-1}\alpha^{-1/2}$ \\ 
        \bottomrule
        \end{tabular}}
	\fi
	\caption{\small BIC-optimal $g_{\lambda, 2}(n, \alpha)$ for statistics $D_n$, $W^2_n$, and $A^2_n$, obtained from weighted least squares and forward-backward model selection with scope \eqref{eq:glambdamu} and $\lambda=2, \ldots, 5$. The following goodness-of-fit measures are presented for each selected model: standard deviation $\hat{\sigma}$ of the residuals of upper-tail observations (i.e., $\{\hat\varepsilon_j \mid \alpha_j\leq0.25\}_{j=1}^{J}$), $\operatorname{BIC}$, and $R^2_{\mathrm{adj}}$. In addition, the $\operatorname{MVIF}$ and $\mathrm{pred}_{\operatorname{MVIF}}$, the predictor that takes the maximum variance inflation factor, inform on the multicollinearity of the selected model.}
	\label{table:powers-n}
\end{table}

\begin{table}[ht!]
	\iffigstabs
	\centering
	\small
	\scalebox{1}{
        \begin{tabular}{ m{0.4cm} m{0.2cm} wr{2cm} wr{1.3cm} wr{1cm} wr{1cm} wl{2cm}} 
        \toprule
        $T_n$ & $\mu$ & $\hat{\sigma}$ & $\operatorname{BIC}$ & $R^2_{\mathrm{adj}}$ & $\operatorname{MVIF}$ & $\mathrm{pred}_{\operatorname{MVIF}}$\\ 
        \midrule
        \multirow{4}{0.6cm}{$D_n$} & $2$ & $7.57\cdot10^{-4}$ & $-397236$ & $0.9993$ & $1\cdot10^2$ & $n^{-1/2}\alpha^{-1/2}$ \\ 
        & $3$ & $5.22\cdot10^{-4}$ & $-421343$ & $0.9996$ & $2\cdot10^3$ & $n^{-1/2}\alpha^{-1}$ \\ 
        & $4$ & $3.68\cdot10^{-4}$ & $-444816$ & $0.9998$ & $5\cdot10^4$ & $n^{-1/2}\alpha^{-3/2}$ \\  
        & $5$ & $2.83\cdot10^{-4}$ & $-462192$ & $0.9999$ & $2\cdot10^6$ & $n^{-1}\alpha^{-2}$ \\  
        \hline
        \multirow{4}{0.6cm}{$W^2_n$} & $2$ & $9.06\cdot10^{-4}$ & $-369813$ & $0.9835$ & $1\cdot10^2$ & $n^{-1/2}\alpha^{-1/2}$ \\ 
        & $3$ & $5.85\cdot10^{-4}$ & $-411594$ & $0.9949$ & $2\cdot10^3$ & $n^{-1/2}\alpha^{-1}$ \\ 
        & $4$ & $5.13\cdot10^{-4}$ & $-427884$ & $0.9968$ & $5\cdot10^4$ & $n^{-1}\alpha^{-3/2}$ \\  
        & $5$ & $4.94\cdot10^{-4}$ & $-433359$ & $0.9972$ & $4\cdot10^5$ & $n^{-1}\alpha^{-2}$ \\  
        \hline
        \multirow{4}{0.6cm}{$A^2_n$} & $2$ & $8.11\cdot10^{-4}$ & $-376072$ & $0.8722$ & $2\cdot10^1$ & $n^{-1}\alpha^{-1/2}$ \\ 
        & $3$ & $5.99\cdot10^{-4}$ & $-403690$ & $0.9414$ & $2\cdot10^3$ & $n^{-1/2}\alpha^{-1}$ \\ 
        & $4$ & $4.67\cdot10^{-4}$ & $-430707$ & $0.9726$ & $7\cdot10^3$ & $n^{-1}\alpha^{-3/2}$ \\  
        & $5$ & $4.15\cdot10^{-4}$ & $-447192$ & $0.9828$ & $3\cdot10^5$ & $n^{-1}\alpha^{-2}$ \\  
        \bottomrule
        \end{tabular}}
	\fi
	\caption{\small BIC-optimal $g_{2, \mu}(n, \alpha)$ for statistics $D_n$, $W^2_n$, and $A^2_n$, obtained from weighted least squares and forward-backward model selection with scope \eqref{eq:glambdamu} and $\mu=2, \ldots, 5$. The table entries are analogous to those of Table \ref{table:powers-n}.}
	\label{table:powers-alpha}
\end{table}


Table \ref{table:powers-n} shows that, when increasing $\lambda$ the $\operatorname{BIC}$ decreases marginally. The standard deviation of the residuals $\hat{\sigma}$ only decreases in the sixth decimal, while the $R^2_{\mathrm{adj}}$ remains almost constant for $D_n$, and $W^2_n$. Moreover, the multicollinearity present in the model increases by an order of magnitude per unit increment in $\lambda$, with high values of $\operatorname{MVIF}$. In the case of $A_n^2$, the BIC-optimal model equals that with $\lambda=2$ --- including more powers of $n$ does not improve it.

Regarding the influence of the powers of $\alpha$, Table \ref{table:powers-alpha} shows similar patterns to those in Table \ref{table:powers-n} for the three statistics $D_n$, $W^2_n$, and $A^2_n$. For $D_n$ and $W_n^2$, when including more powers of $\alpha$, the $\operatorname{BIC}$ and $\hat{\sigma}$ decrease marginally, and $R^2_{\mathrm{adj}}$ increases marginally. In exchange, the multicollinearity increases by an order of magnitude per unit increment in $\mu$, also attaining high values of $\operatorname{MVIF}$. For $A^2_n$ the situation is somehow different: from $\mu=2$ to $\mu=4$ there is a significant increase in $R^2_{\mathrm{adj}}$, yet at expenses of a $\times 100$ increase in $\operatorname{MVIF}$ and a more convoluted model.

From the analyzed test statistics, the final form of the saturated model is chosen to be $g_{2, 2}(n, \alpha)$ due to the general small increase in the goodness-of-fit metrics for more complex models and the increment in multicollinearity when increasing $\lambda$ and $\mu$. Importantly, the choice of $g_{2, 2}(n, \alpha)$ allows having a single generic approach for any statistic and provides parsimonious stabilizing transformations.

\section{Selection of the weight function}
\label{supplementary:weights}

Model \myeqref{eq:saturated_g_n_alpha}{6} is estimated through weighted least squares using the samples $\{(n_j,\alpha_j,Y_j)\}_{j=1}^J$, $Y_j:=T_{\infty; \alpha_j}/T_{n_j; \alpha_j}$. The weights considered in \vspace*{-0.05cm}Section \myref{section:n-alpha-mod}{2.2} are $w_j:=n_j^{-1/2}1_{\{0<\alpha_j\leq 0.25\}}$. The term $\smash{n_j^{-1/2}}$ gives heavier weight to the approximation error on lower sample sizes, while the indicator $1_{\{0<\alpha_j\leq 0.25\}}$ reflects the interest in prioritizing the stabilization of the upper tail of the test statistic $T_n$, where accuracy on the determination of exact-$n$ quantiles is crucial for a precise test decision in the standard significance levels.

The effect of adding other factors to the weights is investigated in this section. First, instead of considering a hard thresholding by $1_{\{0<\alpha_j\leq 0.25\}}$, the factor $\smash{\alpha_j^{-1/2}}$ can be introduced to place more weight on the upper quantiles while still incorporating the remaining quantiles. Second, in order to mitigate the heteroscedasticity of the observations, the asymptotic variance of $Y_j$, 
\begin{align}
    \operatorname{A\mathbb{V}ar}[Y_j]=\frac{T_{\infty;\alpha_j}^2\alpha_j\left(1-\alpha_j\right)}{M \cdot T_{n_j;\alpha_j}^4 \cdot \left(f_n(T_{n_j;\alpha_j})\right)^2},\label{eq:avar}
\end{align}
where $M$ is the number of Monte Carlo samples and $f_n$ is the density of $T_n$, can be incorporated to give more weight to ratios with smaller variances. Expression \eqref{eq:avar} is obtained with the delta method from the standard errors of sample quantiles \citep[see, e.g.,][Section 2.3.3]{Serfling1980|SM}. The evaluation of $f_n$ can be done by differentiating a monotone spline interpolation of its cdf based on the saved quantiles $\{T_{n_j;\alpha_j}: n_j\in\mathcal{N},\,\alpha_j\in\mathcal{A}\}$ (see Section \myref{section:n-alpha-mod}{2.2}).

Different combinations of the previous factors shape the following candidates for weight functions:
\begin{itemize}
    \item $w_{1,j}(\alpha_j, Y_j):=1_{\{0<\alpha_j\leq 0.25\}}$.
    \item $w_{2,j}(n_j, \alpha_j):=n_j^{-1/2}1_{\{0<\alpha_j\leq 0.25\}}$.
    \item $w_{3,j}(\alpha_j, Y_j):=\operatorname{A\mathbb{V}ar}^{-1/2}[Y_j]1_{\{0<\alpha_j\leq 0.25\}}$.
    \item $w_{4,j}(n_j, \alpha_j, Y_j):=n_j^{-1/2}\operatorname{A\mathbb{V}ar}^{-1/2}[Y_j]1_{\{0<\alpha_j\leq 0.25\}}$.
    \item $w_{5,j}(\alpha_j, Y_j):=\operatorname{A\mathbb{V}ar}^{-1/2}[Y_{j}]$.
    \item $w_{6,j}(n_j, \alpha_j, Y_j):=n_j^{-1/2}\operatorname{A\mathbb{V}ar}^{-1/2}[Y_j]$.
    \item $w_{7,j}(n_j, \alpha_j):=(n_j\cdot\alpha_j)^{-1/2}$.
\end{itemize}

The following analysis compares the results of the weighted least squares regression on model \myeqref{eq:saturated_g_n_alpha}{6} with weights computed by $w_{k,j}$, $k=1,\ldots,7$, according to the methodology described in Section \myref{section:n-alpha-mod}{2.2}, except for the last dropping step. The standard deviation of the residuals for the statistics $D_n$, $W^2_n$, and $A^2_n$ is presented from two perspectives: Table \ref{table:weights-alpha} shows the residuals based on the $\alpha_j$ value, while in Table \ref{table:weights-n} the residuals are divided according to the sample size $n_j$. In addition, Table \ref{table:weights-n-alpha0.25} shows the residuals depending on $n_j$ only for upper-tail observations.

\begin{table}[ht!]
	\iffigstabs
	\centering
	\small
        \scalebox{1}{
        \begin{tabular}{m{0.35cm} m{0.5cm} wr{2cm} wr{2cm} wr{2cm} wr{2cm}}
        \toprule
        \multirow{2}{0.4cm}{$w_k$} & \multirow{2}{0.4cm}{$T_n$} & \multicolumn{4}{c}{Standard deviation} \\
        & & $\alpha\in(0,0.25]$ & $\alpha\in(0.25,0.5]$ & $\alpha\in(0.5,0.75]$ & $\alpha\in(0.75,1)$ \\ 
        \midrule
        \multirow{3}{0.6cm}{$w_1$} & $D_n$ & $\mathbf{7.55\cdot10^{-4}}$ & $1.66\cdot10^{-3}$ & $3.11\cdot10^{-3}$ & $7.24\cdot10^{-3}$ \\
        & $W^2_n$ & $\mathbf{9.03\cdot10^{-4}}$ & $1.90\cdot10^{-3}$ & $4.08\cdot10^{-3}$ & $1.01\cdot10^{-2}$ \\
        & $A^2_n$ & $\mathbf{8.11\cdot10^{-4}}$ & $1.50\cdot10^{-3}$ & $2.13\cdot10^{-3}$ & $2.76\cdot10^{-3}$ \\
        \hline
        \multirow{3}{0.6cm}{$w_2$} & $D_n$ & $7.57\cdot10^{-4}$ & $1.67\cdot10^{-3}$ & $3.13\cdot10^{-3}$ & $7.25\cdot10^{-3}$ \\
        & $W^2_n$ & $9.06\cdot10^{-4}$ & $1.90\cdot10^{-3}$ & $4.07\cdot10^{-3}$ & $1.01\cdot10^{-2}$ \\
        & $A^2_n$ & $\mathbf{8.11\cdot10^{-4}}$ & $1.50\cdot10^{-3}$ & $2.13\cdot10^{-3}$ & $2.76\cdot10^{-3}$ \\
        \hline
        \multirow{3}{0.6cm}{$w_3$} & $D_n$ & $8.03\cdot10^{-4}$ & $1.60\cdot10^{-3}$ & $3.05\cdot10^{-3}$ & $7.18\cdot10^{-3}$ \\
        & $W^2_n$ & $9.43\cdot10^{-4}$ & $1.71\cdot10^{-3}$ & $3.83\cdot10^{-3}$ & $9.84\cdot10^{-3}$ \\
        & $A^2_n$ & $8.53\cdot10^{-4}$ & $1.30\cdot10^{-3}$ & $1.88\cdot10^{-3}$ & $2.69\cdot10^{-3}$ \\
        \hline
        \multirow{3}{0.6cm}{$w_4$} & $D_n$ & $8.10\cdot10^{-4}$ & $1.61\cdot10^{-3}$ & $3.06\cdot10^{-3}$ & $7.19\cdot10^{-3}$ \\
        & $W^2_n$ & $9.46\cdot10^{-4}$ & $1.71\cdot10^{-3}$ & $3.83\cdot10^{-3}$ & $9.83\cdot10^{-3}$ \\
        & $A^2_n$ & $8.54\cdot10^{-4}$ & $1.30\cdot10^{-3}$ & $1.87\cdot10^{-3}$ & $2.69\cdot10^{-3}$ \\
        \hline
        \multirow{3}{0.6cm}{$w_5$} & $D_n$ & $6.26\cdot10^{-3}$ & $\mathbf{1.27\cdot10^{-3}}$ & $\mathbf{1.24\cdot10^{-3}}$ & $\mathbf{5.07\cdot10^{-3}}$ \\
        & $W^2_n$ & $4.45\cdot10^{-3}$ & $2.07\cdot10^{-3}$ & $1.17\cdot10^{-3}$ & $\mathbf{5.79\cdot10^{-3}}$ \\
        & $A^2_n$ & $1.47\cdot10^{-3}$ & $\mathbf{3.73\cdot10^{-4}}$ & $\mathbf{6.42\cdot10^{-4}}$ & $2.73\cdot10^{-3}$ \\
        \hline
        \multirow{3}{0.6cm}{$w_6$} & $D_n$ & $6.27\cdot10^{-3}$ & $\mathbf{1.27\cdot10^{-3}}$ & $\mathbf{1.24\cdot10^{-3}}$ & $5.08\cdot10^{-3}$ \\
        & $W^2_n$ & $4.45\cdot10^{-3}$ & $2.05\cdot10^{-3}$ & $\mathbf{1.16\cdot10^{-3}}$ & $5.80\cdot10^{-3}$ \\
        & $A^2_n$ & $1.48\cdot10^{-3}$ & $\mathbf{3.73\cdot10^{-4}}$ & $6.46\cdot10^{-4}$ & $2.73\cdot10^{-3}$ \\
        \hline
        \multirow{3}{0.6cm}{$w_7$} & $D_n$ & $2.52\cdot10^{-3}$ & $1.28\cdot10^{-3}$ & $1.55\cdot10^{-3}$ & $5.66\cdot10^{-3}$ \\
        & $W^2_n$ & $2.42\cdot10^{-3}$ & $\mathbf{1.39\cdot10^{-3}}$ & $1.24\cdot10^{-3}$ & $7.21\cdot10^{-3}$ \\
        & $A^2_n$ & $1.15\cdot10^{-3}$ & $6.36\cdot10^{-4}$ & $1.14\cdot10^{-3}$ & $\mathbf{2.61\cdot10^{-3}}$ \\
        \bottomrule
        \end{tabular}}
	\fi
	\caption{\small Standard deviation of residuals $\{\hat\varepsilon_j\}_{j=1}^{J}$ of the weighted least squares regression of \myeqref{eq:saturated_g_n_alpha}{6} with weights $\{w_{k,j}\}_{j=1}^{J}$, $k=1,\ldots,7$, for statistics $D_n$, $W^2_n$, and $A^2_n$. Residuals are presented in four blocks, each one considering the residuals of the observations whose $\alpha$ value lies within the interval in the column header. Bold highlights the best-performing weight per statistic and $\alpha$-block.}
	\label{table:weights-alpha}
\end{table}


First, we analyze weights $w_1$, $w_2$, $w_3$, and $w_4$, all of which have the factor $1_{\{0<\alpha_j\leq 0.25\}}$ in common. As expected, they present the lowest errors for the upper tail $\alpha\leq0.25$. Despite presenting higher residual deviation for $\alpha>0.25$, the errors differ only by about $\times 2$ with respect to $w_5$ and $w_6$ in average (Table \ref{table:weights-alpha}). More importantly, $w_1$ and $w_2$ produce the smallest residuals for all sample sizes in the upper tail, $w_2$ exhibiting slightly smaller values for small sample sizes, $n\in[5,10)$ (Table \ref{table:weights-n-alpha0.25}).

\begin{table}[ht!]
	\iffigstabs
	\centering
	\small
	\scalebox{1}{
        \begin{tabular}{m{0.4cm} m{0.5cm} wr{2cm} wr{2cm} wr{2cm}}
        \toprule
        \multirow{2}{0.4cm}{$w_k$} & \multirow{2}{0.4cm}{$T_n$} & \multicolumn{3}{c}{Standard deviation} \\
        & & $n\in[5, 10)$ & $n\in[10, 100)$ & $n\in[100, 300)$ \\ 
        \midrule
        \multirow{3}{0.4cm}{$w_1$} & $D_n$ & $1.28\cdot10^{-2}$ & $7.25\cdot10^{-3}$ & $4.11\cdot10^{-3}$ \\ 
        & $W^2_n$ & $1.96\cdot10^{-2}$ & $4.54\cdot10^{-3}$ & $8.21\cdot10^{-4}$ \\ 
        & $A^2_n$ & $7.40\cdot10^{-3}$ & $1.43\cdot10^{-3}$ & $4.19\cdot10^{-4}$  \\ 
        \hline
        \multirow{3}{0.4cm}{$w_2$} & $D_n$ & $1.28\cdot10^{-2}$ & $7.24\cdot10^{-3}$ & $4.10\cdot10^{-3}$ \\ 
        & $W^2_n$ & $1.96\cdot10^{-2}$ & $4.55\cdot10^{-3}$ & $8.37\cdot10^{-4}$ \\ 
        & $A^2_n$ & $7.40\cdot10^{-3}$ & $1.43\cdot10^{-3}$ & $4.19\cdot10^{-4}$ \\ 
        \hline
        \multirow{3}{0.4cm}{$w_3$} & $D_n$ & $1.26\cdot10^{-2}$ & $7.13\cdot10^{-3}$ & $4.03\cdot10^{-3}$ \\ 
        & $W^2_n$ & $1.92\cdot10^{-2}$ & $4.41\cdot10^{-3}$ & $8.03\cdot10^{-4}$  \\ 
        & $A^2_n$ & $7.19\cdot10^{-3}$ & $1.32\cdot10^{-3}$ & $4.07\cdot10^{-4}$  \\ 
        \hline
        \multirow{3}{0.4cm}{$w_4$} & $D_n$ & $1.27\cdot10^{-2}$ & $7.12\cdot10^{-3}$ & $4.02\cdot10^{-3}$ \\ 
        & $W^2_n$ & $1.92\cdot10^{-2}$ & $4.41\cdot10^{-3}$ & $8.17\cdot10^{-4}$  \\ 
        & $A^2_n$ & $7.19\cdot10^{-3}$ & $1.32\cdot10^{-3}$ & $4.07\cdot10^{-4}$ \\ 
        \hline
        \multirow{3}{0.4cm}{$w_5$} & $D_n$ & $\mathbf{1.07\cdot10^{-2}}$ & $\mathbf{5.84\cdot10^{-3}}$ & $\mathbf{3.44\cdot10^{-3}}$ \\ 
        & $W^2_n$ & $\mathbf{1.66\cdot10^{-2}}$ & $\mathbf{3.34\cdot10^{-3}}$ & $\mathbf{7.35\cdot10^{-4}}$ \\ 
        & $A^2_n$ & $7.08\cdot10^{-3}$ & $\mathbf{1.12\cdot10^{-3}}$ & $\mathbf{3.91\cdot10^{-4}}$ \\ 
        \hline
        \multirow{3}{0.4cm}{$w_6$} & $D_n$ & $\mathbf{1.07\cdot10^{-2}}$ & $\mathbf{5.84\cdot10^{-3}}$ & $\mathbf{3.44\cdot10^{-3}}$ \\ 
        & $W^2_n$ & $\mathbf{1.66\cdot10^{-2}}$ & $3.35\cdot10^{-3}$ & $7.39\cdot10^{-4}$ \\ 
        & $A^2_n$ & $\mathbf{7.07\cdot10^{-3}}$ & $\mathbf{1.12\cdot10^{-3}}$ & $3.95\cdot10^{-4}$ \\ 
        \hline
        \multirow{3}{0.4cm}{$w_7$} & $D_n$ & $1.13\cdot10^{-2}$ & $6.23\cdot10^{-3}$ & $3.64\cdot10^{-3}$ \\ 
        & $W^2_n$ & $1.76\cdot10^{-2}$ & $3.59\cdot10^{-3}$ & $7.75\cdot10^{-4}$ \\ 
        & $A^2_n$ & $7.13\cdot10^{-3}$ & $1.16\cdot10^{-3}$ & $\mathbf{3.91\cdot10^{-4}}$ \\ 
        \bottomrule
        \end{tabular}}
	\fi
	\caption{\small Standard deviation of residuals $\{\hat\varepsilon_j\}_{j=1}^{J}$ of the weighted least squares regression of \myeqref{eq:saturated_g_n_alpha}{6} with weights $\{w_{k,j}\}_{j=1}^{J}$, $k=1,\ldots,7$, for statistics $D_n$, $W^2_n$, and $A^2_n$. The results are divided into three blocks, each one considering the residuals of the observations whose $n$ value lies within the interval in the column header. Bold highlights the best-performing weight per statistic and $n$-block.}
	\label{table:weights-n}
\end{table}

Second, $w_5$ and $w_6$ factor in the asymptotic variance, showing both similar results. In particular, they attain the lowest errors for $\alpha>0.25$ (Table \ref{table:weights-alpha}) and small-to-moderate sample sizes (Table \ref{table:weights-n}). However, the standard deviation of upper-tail residuals is one order of magnitude higher than $w_1$, $w_2$, $w_3$, and $w_4$ (Table \ref{table:weights-alpha}).

Finally, $w_7$ weights observations by $\smash{\alpha_j^{-1/2}}$. Its behavior is similar to $w_5$ and $w_6$ for $\alpha>0.25$ and all sample sizes. Yet, errors in the upper tail are lower, but still about $\times3$ higher than $w_1$ and $w_2$.

From the previous observations, the final weight function chosen to fit model \myeqref{eq:saturated_g_n_alpha}{6} is $w_2$ due to two main reasons: (\textit{i}) the significant difference of errors for $\alpha\leq0.25$ against $w_5$, $w_6$, and $w_7$; and (\textit{ii}) the lower residual deviation in the upper tail for small sample sizes when compared to $w_1$, $w_3$, and $w_4$.

\begin{table}[ht!]
	\iffigstabs
	\centering
	\small
	\scalebox{1}{
        \begin{tabular}{m{0.4cm} m{0.5cm} wr{2cm} wr{2cm} wr{2cm}}
        \toprule
        \multirow{2}{0.4cm}{$w_k$} & \multirow{2}{0.4cm}{$T_n$} & \multicolumn{3}{c}{Standard deviation} \\
        & & $n\in[5, 10)$ & $n\in[10, 100)$ & $n\in[100, 300)$ \\ 
        \midrule
        \multirow{3}{0.4cm}{$w_1$} & $D_n$ & $1.21\cdot10^{-3}$ & $\mathbf{8.21\cdot10^{-4}}$ & $\mathbf{5.23\cdot10^{-4}}$ \\ 
        & $W^2_n$ & $3.03\cdot10^{-3}$ & $\mathbf{7.93\cdot10^{-4}}$ & $\mathbf{5.09\cdot10^{-4}}$ \\ 
        & $A^2_n$ & $\mathbf{2.84\cdot10^{-3}}$ & $\mathbf{6.96\cdot10^{-4}}$ & $\mathbf{4.41\cdot10^{-4}}$ \\ 
        \hline 
        \multirow{3}{0.4cm}{$w_2$} & $D_n$ & $\mathbf{1.19\cdot10^{-3}}$ & $8.22\cdot10^{-4}$ & $5.29\cdot10^{-4}$ \\ 
        & $W^2_n$ & $\mathbf{3.01\cdot10^{-3}}$ & $7.98\cdot10^{-4}$ & $5.25\cdot10^{-4}$ \\ 
        & $A^2_n$ & $\mathbf{2.84\cdot10^{-3}}$ & $\mathbf{6.96\cdot10^{-4}}$ & $\mathbf{4.41\cdot10^{-4}}$ \\ 
        \hline 
        \multirow{3}{0.4cm}{$w_3$} & $D_n$ & $1.26\cdot10^{-3}$ & $8.71\cdot10^{-4}$ & $5.67\cdot10^{-4}$ \\ 
        & $W^2_n$ & $3.19\cdot10^{-3}$ & $8.32\cdot10^{-4}$ & $5.11\cdot10^{-4}$ \\ 
        & $A^2_n$ & $3.02\cdot10^{-3}$ & $7.33\cdot10^{-4}$ & $4.42\cdot10^{-4}$ \\ 
        \hline 
        \multirow{3}{0.4cm}{$w_4$} & $D_n$ & $1.23\cdot10^{-3}$ & $8.80\cdot10^{-4}$ & $5.80\cdot10^{-4}$ \\ 
        & $W^2_n$ & $3.17\cdot10^{-3}$ & $8.35\cdot10^{-4}$ & $5.26\cdot10^{-4}$ \\ 
        & $A^2_n$ & $3.02\cdot10^{-3}$ & $7.33\cdot10^{-4}$ & $4.43\cdot10^{-4}$ \\ 
        \hline 
        \multirow{3}{0.4cm}{$w_5$} & $D_n$ & $1.18\cdot10^{-2}$ & $6.74\cdot10^{-3}$ & $4.05\cdot10^{-3}$ \\ 
        & $W^2_n$ & $1.78\cdot10^{-2}$ & $3.63\cdot10^{-3}$ & $7.71\cdot10^{-4}$ \\ 
        & $A^2_n$ & $5.24\cdot10^{-3}$ & $1.26\cdot10^{-3}$ & $5.15\cdot10^{-4}$ \\ 
        \hline 
        \multirow{3}{0.4cm}{$w_6$} & $D_n$ & $1.17\cdot10^{-2}$ & $6.76\cdot10^{-3}$ & $4.07\cdot10^{-3}$ \\ 
        & $W^2_n$ & $1.79\cdot10^{-2}$ & $3.63\cdot10^{-3}$ & $7.77\cdot10^{-4}$ \\ 
        & $A^2_n$ & $5.19\cdot10^{-3}$ & $1.27\cdot10^{-3}$ & $5.27\cdot10^{-4}$ \\ 
        \hline 
        \multirow{3}{0.4cm}{$w_7$} & $D_n$ & $4.34\cdot10^{-3}$ & $2.54\cdot10^{-3}$ & $1.47\cdot10^{-3}$ \\ 
        & $W^2_n$ & $6.30\cdot10^{-3}$ & $1.76\cdot10^{-3}$ & $5.32\cdot10^{-4}$ \\ 
        & $A^2_n$ & $3.18\cdot10^{-3}$ & $8.83\cdot10^{-4}$ & $\mathbf{4.41\cdot10^{-4}}$ \\ 
        \bottomrule
        \end{tabular}}
	\fi
	\caption{\small Standard deviation of upper-tail residuals $\{\hat\varepsilon_j \mid \alpha_j\leq0.25\}_{j=1}^{J}$ of the weighted least squares regression of \myeqref{eq:saturated_g_n_alpha}{6} with weights $\{w_{k,j}\}_{j=1}^{J}$, $k=1,\ldots,7$, for statistics $D_n$, $W^2_n$, and $A^2_n$. Residuals are presented in three blocks, each one considering the residuals of the observations whose $n$ value lies within the interval in the column header. Bold highlights the best-performing weight per statistic and $n$-block.}
	\label{table:weights-n-alpha0.25}
\end{table}

\fi


\begin{thebibliography}{}
	
	\bibitem[Agostinelli and Lund, 2017]{Agostinelli2017}
	Agostinelli, C. and Lund, U. (2017).
	\newblock {\em {R} package {circular}: Circular Statistics}.
	\newblock {R} package version 0.4-93.
	
	\bibitem[Arsham, 1988]{Arsham1988}
	Arsham, H. (1988).
	\newblock Kuiper's {P}-value as a measuring tool and decision procedure for the
	goodness-of-fit test.
	\newblock {\em J. Appl. Stat.}, 15(2):131--135.
	
	\bibitem[Babcock, 1961]{Babcock1961}
	Babcock, H.~W. (1961).
	\newblock The topology of the {S}un's magnetic field and the 22-year cycle.
	\newblock {\em Astrophys. J.}, 133(2):572--587.
	
	\bibitem[Bakshaev, 2010]{Bakshaev2010}
	Bakshaev, A. (2010).
	\newblock {$N$}-distance tests of uniformity on the hypersphere.
	\newblock {\em Nonlinear Anal. Model. Control.}, 15(1):15--8.
	
	\bibitem[Baranyi et~al., 2016]{Baranyi2016}
	Baranyi, T., Gy\H{o}ri, L., and Ludm\'any, A. (2016).
	\newblock On-line tools for solar data compiled at the {D}ebrecen observatory
	and their extensions with the {G}reenwich sunspot data.
	\newblock {\em Sol. Phys.}, 291(9):3081--3102.
	
	\bibitem[Benjamini and Yekutieli, 2001]{Benjamini2001}
	Benjamini, Y. and Yekutieli, D. (2001).
	\newblock The control of the false discovery rate in multiple testing under
	dependency.
	\newblock {\em Ann. Stat.}, 29(4):1165--1188.
	
	\bibitem[Berdyugina and Usoskin, 2003]{Berdyugina2003}
	Berdyugina, S.~V. and Usoskin, I.~G. (2003).
	\newblock Active longitudes in sunspot activity: Century scale persistence.
	\newblock {\em Astron. Astrophys.}, 405(3):1121--1128.
	
	\bibitem[Birnbaum, 1952]{Birnbaum1952}
	Birnbaum, Z.~W. (1952).
	\newblock Numerical tabulation of the distribution of {K}olmogorov's statistic
	for finite sample size.
	\newblock {\em J. Am. Stat. Assoc.}, 47(259):425--441.
	
	\bibitem[Bogart, 1982]{Bogart1982}
	Bogart, R.~S. (1982).
	\newblock Recurrence of solar activity: Evidence for active longitudes.
	\newblock {\em Solar Phys.}, 76(1):155--165.
	
	\bibitem[Brown and Harvey, 2007]{Brown2007}
	Brown, J.~R. and Harvey, M.~E. (2007).
	\newblock Rational arithmetic mathematica functions to evaluate the one-sided
	one sample {K}–{S} cumulative sampling distribution.
	\newblock {\em J. Stat. Softw.}, 19(6):1--32.
	
	\bibitem[Crown, 2000]{Crown2000}
	Crown, J.~S. (2000).
	\newblock Percentage points for directional {A}nderson--{D}arling
	goodness-of-fit tests.
	\newblock {\em Commun. Stat. Simul. Comput.}, 29(2):523--532.
	
	\bibitem[Cs\"org\"o and Faraway, 1996]{Csorgo1996}
	Cs\"org\"o, S. and Faraway, J.~J. (1996).
	\newblock The exact and asymptotic distributions of {C}ram\'er-von {M}ises
	statistics.
	\newblock {\em J. R. Stat. Soc. Ser. B Methodol.}, 58(1):221--234.
	
	\bibitem[Cuesta-Albertos et~al., 2009]{Cuesta-Albertos2009}
	Cuesta-Albertos, J.~A., Cuevas, A., and Fraiman, R. (2009).
	\newblock On projection-based tests for directional and compositional data.
	\newblock {\em Stat. Comput.}, 19(4):367--380.
	
	\bibitem[D'Agostino and Stephens, 1986]{D'Agostino1986}
	D'Agostino, R.~B. and Stephens, M.~A. (1986).
	\newblock {\em Goodness-of-Fit Techniques}, volume~68 of {\em Statistics:
		Textbooks and Monographs}.
	\newblock Marcel Dekker, New York.
	
	\bibitem[Di~Marzio et~al., 2012]{DiMarzio2012}
	Di~Marzio, M., Panzera, A., and Taylor, C.~C. (2012).
	\newblock Smooth estimation of circular cumulative distribution functions and
	quantiles.
	\newblock {\em J. Nonparametr. Stat.}, 24(4):935--949.
	
	\bibitem[Dufour and Maag, 1978]{Dufour1978}
	Dufour, R. and Maag, U.~R. (1978).
	\newblock Distribution results for modified {K}olmogorov--{S}mirnov statistics
	for truncated or censored.
	\newblock {\em Technometrics}, 20(1):29--32.
	
	\bibitem[Durbin, 1969]{Durbin1969}
	Durbin, J. (1969).
	\newblock Tests for serial correlation in regression analysis based on the
	periodogram of least-squares residuals.
	\newblock {\em Biometrika}, 56(1):1--15.
	
	\bibitem[Durbin, 1973]{Durbin1973}
	Durbin, J. (1973).
	\newblock {\em Distribution Theory for Tests Based on the Sample Distribution
		Function}, volume~9 of {\em CBMS-NSF Regional Conference Series in Applied
		Mathematics}.
	\newblock Society for Industrial and Applied Mathematics, Philadelphia.
	
	\bibitem[Eddelbuettel and Fran\c{c}ois, 2011]{Eddelbuettel2011}
	Eddelbuettel, D. and Fran\c{c}ois, R. (2011).
	\newblock {Rcpp}: Seamless {R} and {C++} integration.
	\newblock {\em J. Stat. Softw.}, 40(8):1--18.
	
	\bibitem[Elstner and Korhonen, 2005]{Elstner2005}
	Elstner, D. and Korhonen, H. (2005).
	\newblock Flip-flop phenomenon: observations and theory.
	\newblock {\em Astron. Nachr.}, 326(3-4):278--282.
	
	\bibitem[Facchinetti, 2009]{Facchinetti2009}
	Facchinetti, S. (2009).
	\newblock A procedure to find exact critical values of {K}olmogorov--{S}mirnov
	test.
	\newblock {\em Stat. App.}, 21(3--4):337--359.
	
	\bibitem[Faraway et~al., 2019]{Faraway2019}
	Faraway, J., Marsaglia, G., Marsaglia, J., and Baddeley, A. (2019).
	\newblock {\em {goftest}: Classical Goodness-of-Fit Tests for Univariate
		Distributions}.
	\newblock {R} package version 1.2-2.
	
	\bibitem[Garc\'ia-Portugu\'es, 2013]{Garcia-Portugues2013a}
	Garc\'ia-Portugu\'es, E. (2013).
	\newblock Exact risk improvement of bandwidth selectors for kernel density
	estimation with directional data.
	\newblock {\em Electron. J. Stat.}, 7:1655--1685.
	
	\bibitem[Garc\'ia-Portugu\'es et~al., 2013]{Garcia-Portugues2013b}
	Garc\'ia-Portugu\'es, E., Crujeiras, R.~M., and Gonz\'alez-Manteiga, W. (2013).
	\newblock Kernel density estimation for directional-linear data.
	\newblock {\em J. Multivar. Anal.}, 121:152--175.
	
	\bibitem[Garc\'ia-Portugu\'es et~al., 2020]{Garcia-Portugues2020b}
	Garc\'ia-Portugu\'es, E., Navarro-Esteban, P., and Cuesta-Albertos, J.~A.
	(2020).
	\newblock On a projection-based class of uniformity tests on the hypersphere.
	\newblock {\em arXiv:2008.09897}.
	
	\bibitem[Garc\'ia-Portugu\'es et~al., 2021]{Garcia-Portugues2020e}
	Garc\'ia-Portugu\'es, E., Paindaveine, D., and Verdebout, T. (2021).
	\newblock {\em {rotasym}: Tests for Rotational Symmetry on the Hypersphere}.
	\newblock {R} package version 1.1.3.
	
	\bibitem[Garc\'ia-Portugu\'es and Verdebout, 2018]{Garcia-Portugues2020a}
	Garc\'ia-Portugu\'es, E. and Verdebout, T. (2018).
	\newblock A review of uniformity tests on the hypersphere.
	\newblock {\em arXiv:1804.00286}.
	
	\bibitem[Garc\'ia-Portugu\'es and Verdebout, 2021]{Garcia-Portugues2020c}
	Garc\'ia-Portugu\'es, E. and Verdebout, T. (2021).
	\newblock {\em {sphunif}: Uniformity Tests on the Circle, Sphere, and
		Hypersphere}.
	\newblock {R} package version 1.0.1.
	
	\bibitem[Gy\H{o}ri et~al., 2016]{Gyoeri2016}
	Gy\H{o}ri, L., Baranyi, T., and Lud\'amny, A. (2016).
	\newblock Comparative analysis of {D}ebrecen sunspot catalogues.
	\newblock {\em Mon. Not. R. Astron. Soc.}, 465(2):1259--1273.
	
	\bibitem[Hegazy and Green, 1975]{Hegazy1975}
	Hegazy, Y. A.~S. and Green, J.~R. (1975).
	\newblock Some new goodness-of-fit tests using order statistics.
	\newblock {\em J. R. Stat. Soc. Ser. C Appl. Stat.}, 24(3):299--308.
	
	\bibitem[Heo et~al., 2013]{Heo2013}
	Heo, J.-H., Shin, H., Nam, W., Om, J., and Jeong, C. (2013).
	\newblock Approximation of modified {A}nderson--{D}arling test statistics for
	extreme value distributions with unknown shape parameter.
	\newblock {\em J. Hydrol.}, 499:41--49.
	
	\bibitem[Johannes and Rasche, 1980]{Johannes1980}
	Johannes, J.~M. and Rasche, R.~H. (1980).
	\newblock Additional information on significance values for {D}urbin's {$c^+$},
	{$c^-$} and {$c$} statistics.
	\newblock {\em Biometrika}, 67(2):511--514.
	
	\bibitem[Jupp and Kume, 2020]{Jupp2020}
	Jupp, P.~E. and Kume, A. (2020).
	\newblock Measures of goodness of fit obtained by almost-canonical
	transformations on {R}iemannian manifolds.
	\newblock {\em J. Multivar. Anal.}, 176:104579.
	
	\bibitem[Knott, 1974]{Knott1974}
	Knott, M. (1974).
	\newblock The distribution of the {C}ram\'er--von {M}ises statistic for small
	sample sizes.
	\newblock {\em J. R. Stat. Soc. Ser. B Methodol.}, 36(3):430--438.
	
	\bibitem[Kuiper, 1960]{Kuiper1960}
	Kuiper, N.~H. (1960).
	\newblock Tests concerning random points on the circle.
	\newblock {\em K. Ned. Akad. Van Wet. A}, 63:38--47.
	
	\bibitem[Lewis, 1961]{Lewis1961}
	Lewis, P. A.~W. (1961).
	\newblock Distribution of the {A}nderson--{D}arling statistic.
	\newblock {\em Ann. Math. Stat.}, 32(4):1118--1124.
	
	\bibitem[Maag and Dicaire, 1971]{Maag1971}
	Maag, U.~R. and Dicaire, G. (1971).
	\newblock On {K}olmogorov--{S}mirnov type one-sample statistics.
	\newblock {\em Biometrika}, 58(3):653--656.
	
	\bibitem[Mardia and Jupp, 1999]{Mardia1999a}
	Mardia, K.~V. and Jupp, P.~E. (1999).
	\newblock {\em Directional Statistics}.
	\newblock Wiley Series in Probability and Statistics. Wiley, Chichester.
	
	\bibitem[Marks, 1998]{Marks1998}
	Marks, N.~B. (1998).
	\newblock Modification of the {K}olmogorov--{S}mirnov test for the {E}rlang-2
	distribution.
	\newblock {\em Commun. Stat. Simul. Comput.}, 27(1):39--49.
	
	\bibitem[Marks, 2007]{Marks2007}
	Marks, N.~B. (2007).
	\newblock {K}olmogorov--{S}mirnov test statistic and critical values for the
	{E}rlang-3 and {E}rlang-4 distributions.
	\newblock {\em J. Appl. Stat.}, 34(8):899--906.
	
	\bibitem[Marsaglia and Marsaglia, 2004]{Marsaglia2004}
	Marsaglia, G. and Marsaglia, J. (2004).
	\newblock Evaluating the {A}nderson--{D}arling distribution.
	\newblock {\em J. Stat. Softw.}, 9(2):1--5.
	
	\bibitem[Marsaglia et~al., 2003]{Marsaglia2003}
	Marsaglia, G., Tsang, W.~W., and Wang, J. (2003).
	\newblock Evaluating {K}olmogorov's distribution.
	\newblock {\em J. Stat. Softw.}, 8(18):1--4.
	
	\bibitem[Marshall, 1958]{Marshall1958}
	Marshall, A.~W. (1958).
	\newblock The small sample distribution of {$n\omega^2_n$}.
	\newblock {\em Ann. Math. Stat.}, 29(1):307--309.
	
	\bibitem[Massey, 1950]{Massey1950}
	Massey, F.~J. (1950).
	\newblock A note on the estimation of a distribution function by confidence
	limits.
	\newblock {\em Ann. Stat.}, 21(1):116--119.
	
	\bibitem[Massey, 1951]{Massey1951}
	Massey, F.~J. (1951).
	\newblock The {K}olmogorov--{S}mirnov test for goodness of fit.
	\newblock {\em J. Am. Stat. Assoc.}, 46(253):68--78.
	
	\bibitem[Mersmann, 2019]{Mersmann2019}
	Mersmann, O. (2019).
	\newblock {\em microbenchmark: Accurate Timing Functions}.
	\newblock R package version 1.4-7.
	
	\bibitem[Millard, 2013]{Millard2013}
	Millard, S.~P. (2013).
	\newblock {\em {EnvStats}}.
	\newblock Springer, New York.
	
	\bibitem[Pearson and Stephens, 1962]{Pearson1962}
	Pearson, E.~S. and Stephens, M.~A. (1962).
	\newblock The goodness-of-fit tests based on {$W^2_N$} and {$U^2_N$}.
	\newblock {\em Biometrika}, 49(3/4):397--402.
	
	\bibitem[Pelt et~al., 2010]{Pelt2010}
	Pelt, J., Korpi, M.~J., and Tuominen, I. (2010).
	\newblock Solar active regions: A nonparametric statistical analysis.
	\newblock {\em Astron. Astrophys.}, 513:A48.
	
	\bibitem[Pettitt, 1977]{Pettitt1977}
	Pettitt, A.~N. (1977).
	\newblock Testing the normality of several independent samples using the
	{A}nderson--{D}arling statistic.
	\newblock {\em J. R. Stat. Soc. Ser. C Appl. Stat.}, 26(2):156--161.
	
	\bibitem[Pewsey and Garc\'ia-Portugu\'es, 2021]{Pewsey2021}
	Pewsey, A. and Garc\'ia-Portugu\'es, E. (2021).
	\newblock Recent advances in directional statistics.
	\newblock {\em Test}, 30(1):1--58.
	
	\bibitem[Quesenberry and Miller~Jr, 1977]{Quesenberry1977}
	Quesenberry, C.~P. and Miller~Jr, F.~L. (1977).
	\newblock Power studies of some tests for uniformity.
	\newblock {\em J. Stat. Comput. Simul.}, 5(3):169--191.
	
	\bibitem[{R Core Team}, 2021]{RCoreTeam2021}
	{R Core Team} (2021).
	\newblock {\em {R}: A Language and Environment for Statistical Computing}.
	\newblock R Foundation for Statistical Computing, Vienna.
	
	\bibitem[Serfling, 1980]{Serfling1980}
	Serfling, R.~J. (1980).
	\newblock {\em Approximation Theorems of Mathematical Statistics}.
	\newblock Wiley Series in Probability and Statistics. Wiley, New York.
	
	\bibitem[Silverman, 1986]{Silverman1986}
	Silverman, B.~W. (1986).
	\newblock {\em Density Estimation for Statistics and Data Analysis}.
	\newblock Monographs on Statistics and Applied Probability. Chapman \& Hall,
	London.
	
	\bibitem[Stephens, 1965]{Stephens1965}
	Stephens, M.~A. (1965).
	\newblock The goodness-of-fit statistic {$V_n$}: distribution and significance
	points.
	\newblock {\em Biometrika}, 52(3/4):309--321.
	
	\bibitem[Stephens, 1970]{Stephens1970}
	Stephens, M.~A. (1970).
	\newblock Use of the {K}olmogorov--{S}mirnov, {C}ram\'er-von {M}ises and
	related statistics without extensive tables.
	\newblock {\em J. R. Stat. Soc. Ser. B Methodol.}, 32(1):115--122.
	
	\bibitem[Stephens, 1974]{Stephens1974}
	Stephens, M.~A. (1974).
	\newblock {EDF} statistics for goodness of fit and some comparisons.
	\newblock {\em J. Am. Stat. Assoc.}, 69(347):730--737.
	
	\bibitem[Stephens, 1977]{Stephens1977b}
	Stephens, M.~A. (1977).
	\newblock Goodness of fit for the extreme value distribution.
	\newblock {\em Biometrika}, 64(3):583--588.
	
	\bibitem[Stephens, 1979]{Stephens1979b}
	Stephens, M.~A. (1979).
	\newblock Tests of fit for the logistic distribution based on the empirical
	distribution function.
	\newblock {\em Biometrika}, 66(3):591--595.
	
	\bibitem[Stephens and Maag, 1968]{Stephens1968}
	Stephens, M.~A. and Maag, U.~R. (1968).
	\newblock Further percentage points for {$W^2_{N}$}.
	\newblock {\em Biometrika}, 55(2):428--430.
	
	\bibitem[Tiku, 1965]{Tiku1965}
	Tiku, M.~L. (1965).
	\newblock Chi-square approximations for the distributions of goodness-of-fit
	statistics {$U_n^2$} and {$W_n^2$}.
	\newblock {\em Biometrika}, 52(3/4):630--633.

	\bibitem[de~Toma et~al., 2000]{deToma2000}
	de~Toma, G., White, O.~R., and Harvey, K.~L. (2000).
	\newblock A picture of solar minimum and the onset of solar cycle 23. {I}.
	{G}lobal magnetic field evolution.
	\newblock {\em Astrophys. J.}, 529(2):1101.
	
	\bibitem[Watson, 1961]{Watson1961}
	Watson, G.~S. (1961).
	\newblock Goodness-of-fit tests on a circle.
	\newblock {\em Biometrika}, 48(1/2):109--114.
	
\end{thebibliography}

\begin{thebibliography}{1}
        \bibitem[{Serfling(1980)}]{Serfling1980|SM}
        \bibinfo{author}{Serfling, R.~J.} (\bibinfo{year}{1980}).
        \newblock {\it \bibinfo{title}{Approximation Theorems of Mathematical
          Statistics}\/}.
        \newblock Wiley Series in Probability and Statistics.
        \newblock \bibinfo{address}{New York}: \bibinfo{publisher}{Wiley}.
\end{thebibliography}
\end{document}